\documentclass{bmcart}

\usepackage{fancyhdr}
\usepackage{graphicx}
\usepackage[utf8]{inputenc} 

\usepackage{amssymb}
\usepackage{nicefrac}
\usepackage{hyperref}

\usepackage{verbatim}
\usepackage{etoolbox}
\makeatletter
\preto{\@verbatim}{\topsep=0.5\baselineskip  \partopsep=0.5\baselineskip}
\makeatother

\usepackage{tikz}
\usetikzlibrary{arrows,decorations.pathmorphing,decorations.markings,backgrounds,positioning,fit,calc,shapes,shapes,matrix}
\usepackage{varwidth}

\usepackage{xcolor}
\usepackage{bigfoot}
\usepackage{siunitx}

\newcommand{\mus}[1]{\SI{#1}{\micro \second}}

\definecolor{green}{rgb}{0,1,00}
\definecolor{lightgreen}{rgb}{0,0.5,0}
\definecolor{red}{rgb}{1,0,0}
\definecolor{lightred}{rgb}{0.5,0,0}

\newcounter{inCounter}[section]
\newcommand{\iin}{\refstepcounter{inCounter}{\color{lightgreen}In [{\color{green}\arabic{inCounter}}]:}}
\newcommand{\out}{{\color{lightred}Out[{\color{red}\arabic{inCounter}}]:}}

\newcommand{\ddo}{\textsc{d2o}}
\newcommand{\distributeddataobject}{\emph{distributed\_data\_object}}
\newcommand{\distributeddataobjects}{\emph{distributed\_data\_objects}}
\newcommand{\nifty}{\textsc{NIFTy}}

\definecolor{maroon}{cmyk}{0, 0.87, 0.68, 0.32}
\definecolor{halfgray}{gray}{0.55}
\definecolor{ipython_frame}{RGB}{207, 207, 207}
\definecolor{ipython_bg}{RGB}{247, 247, 247}
\definecolor{ipython_red}{RGB}{186, 33, 33}
\definecolor{ipython_green}{RGB}{0, 128, 0}
\definecolor{ipython_cyan}{RGB}{64, 128, 128}
\definecolor{ipython_purple}{RGB}{170, 34, 255}

\usepackage{listings}
\lstdefinelanguage{iPython}{
    morekeywords={access,and,break,class,continue,def,del,elif,else,except,exec,finally,for,from,global,if,import,in,is,lambda,not,or,pass,print,raise,return,try,while},%
    morekeywords=[2]{abs,all,any,basestring,bin,bool,bytearray,callable,chr,classmethod,cmp,compile,complex,delattr,dict,dir,divmod,enumerate,eval,execfile,file,filter,float,format,frozenset,getattr,globals,hasattr,hash,help,hex,id,input,int,isinstance,issubclass,iter,len,list,locals,long,map,max,memoryview,min,next,object,oct,open,ord,pow,property,range,raw_input,reduce,reload,repr,reversed,round,set,setattr,slice,sorted,staticmethod,str,sum,super,tuple,type,unichr,unicode,vars,xrange,zip,apply,buffer,coerce,intern},%
    sensitive=true,%
    morecomment=[l]\#,%
    morestring=[b]',%
    morestring=[b]",%
    morestring=[s]{'''}{'''},
    morestring=[s]{"""}{"""},
    morestring=[s]{r'}{'},
    morestring=[s]{r"}{"},%
    morestring=[s]{r'''}{'''},%
    morestring=[s]{r"""}{"""},%
    morestring=[s]{u'}{'},
    morestring=[s]{u"}{"},%
    morestring=[s]{u'''}{'''},%
    morestring=[s]{u"""}{"""},%
    literate=
    {á}{{\'a}}1 {é}{{\'e}}1 {í}{{\'i}}1 {ó}{{\'o}}1 {ú}{{\'u}}1
    {Á}{{\'A}}1 {É}{{\'E}}1 {Í}{{\'I}}1 {Ó}{{\'O}}1 {Ú}{{\'U}}1
    {à}{{\`a}}1 {è}{{\`e}}1 {ì}{{\`i}}1 {ò}{{\`o}}1 {ù}{{\`u}}1
    {À}{{\`A}}1 {È}{{\'E}}1 {Ì}{{\`I}}1 {Ò}{{\`O}}1 {Ù}{{\`U}}1
    {ä}{{\"a}}1 {ë}{{\"e}}1 {ï}{{\"i}}1 {ö}{{\"o}}1 {ü}{{\"u}}1
    {Ä}{{\"A}}1 {Ë}{{\"E}}1 {Ï}{{\"I}}1 {Ö}{{\"O}}1 {Ü}{{\"U}}1
    {â}{{\^a}}1 {ê}{{\^e}}1 {î}{{\^i}}1 {ô}{{\^o}}1 {û}{{\^u}}1
    {Â}{{\^A}}1 {Ê}{{\^E}}1 {Î}{{\^I}}1 {Ô}{{\^O}}1 {Û}{{\^U}}1
    {œ}{{\oe}}1 {Œ}{{\OE}}1 {æ}{{\ae}}1 {Æ}{{\AE}}1 {ß}{{\ss}}1
    {ç}{{\c c}}1 {Ç}{{\c C}}1 {ø}{{\o}}1 {å}{{\r a}}1 {Å}{{\r A}}1
    {€}{{\EUR}}1 {£}{{\pounds}}1,
    literate=
    *{+}{{{\color{ipython_purple}+}}}1
    {-}{{{\color{ipython_purple}-}}}1
    {*}{{{\color{ipython_purple}$^\ast$}}}1
    {/}{{{\color{ipython_purple}/}}}1
    {^}{{{\color{ipython_purple}\^{}}}}1
    {?}{{{\color{ipython_purple}?}}}1
    {!}{{{\color{ipython_purple}!}}}1
    {\%}{{{\color{ipython_purple}\%}}}1
    {|}{{{\color{ipython_purple}|}}}1
    {\&}{{{\color{ipython_purple}\&}}}1
    {~}{{{\color{ipython_purple}~}}}1
    {==}{{{\color{ipython_purple}==}}}2
    {<=}{{{\color{ipython_purple}<=}}}2
    {>=}{{{\color{ipython_purple}>=}}}2
    {np.float}{{{np.float}}}7
    {np.complex}{{{np.complex}}}9
    {+=}{{{+=}}}2
    {-=}{{{-=}}}2
    {*=}{{{$^\ast$=}}}2
    {/=}{{{/=}}}2,
    commentstyle=\color{ipython_cyan}\ttfamily,
    stringstyle=\color{ipython_red}\ttfamily,
    keepspaces=true,
    showspaces=false,
    showstringspaces=false,
    rulecolor=\color{black},
    frame=l,
    frameround={t}{t}{t}{t},
    framexleftmargin=0mm,
    numbers=left,
    numberstyle=\tiny\color{black},
    basicstyle=\small\ttfamily,
    keywordstyle=\color{ipython_green}\ttfamily,
    xleftmargin=18pt,
    escapechar=|,escapebegin=\color{ipython_green},
    gobble=4,
    tabsize=4,
}

\usepackage{calc}
\usepackage{pdflscape}
\usepackage{pgfplotstable}
\usepackage{booktabs}
\usepackage{filecontents}

\pgfplotstableset{
    percent cells/.style={
        col sep=comma,
        every head row/.style={before row=\toprule,after row=\midrule},
        every last row/.style={after row=[0.5ex]\bottomrule},
        postproc cell content/.style={
          /pgfplots/table/@cell content/.initial={}{%
              \pgfmathparse{##1<30}
              \ifnum\pgfmathresult=1
                  \pgfmathparse{##1<10}
                  \ifnum\pgfmathresult=1
                      $\mathit{\pgfmathprintnumber[assume math mode=true, fixed, fixed zerofill, precision=2]{##1}\%}$
                  \else
                      $\mathit{\pgfmathprintnumber[assume math mode=true, fixed, fixed zerofill, precision=1]{##1}\%}$
                  \fi
              \else
                  \pgfmathparse{##1<90}
                  \ifnum\pgfmathresult=1
                      \pgfmathprintnumber[assume math mode=true, fixed, fixed zerofill, precision=1]{##1}\%
                  \else
                      $\mathbf{\pgfmathprintnumber[assume math mode=true, fixed, fixed zerofill, precision=1]{##1}\%}$
                  \fi
              \fi
          },
        },
}}

\pgfplotstableset{
    nonpercent cells/.style={
        col sep=comma,
        every head row/.style={before row=\toprule,after row=\midrule},
        every last row/.style={after row=[0.5ex]\bottomrule},
        postproc cell content/.style={
          /pgfplots/table/@cell content/.initial={}{%
            $\mathbf{\pgfmathprintnumber[assume math mode=true, std=-1:1, fixed zerofill, sci zerofill, precision=2]{##1}}$

          },
        },
}}

\startlocaldefs

\renewcommand{\verbatim@font}{\ttfamily\small}

\endlocaldefs

\begin{document}

\begin{frontmatter}

\begin{fmbox}
\dochead{Methodology}


\title{D2O - a distributed data object for parallel high-performance computing in Python}


\author[
   addressref={maxplanck,lmu},                   
   email={theos@mpa-garching.mpg.de}   
]{\inits{TS}\fnm{Theo} \snm{Steininger}}
\author[
addressref={maxplanck,lmu},                   
email={maksim@MPA-Garching.MPG.DE}   
]{\inits{MG}\fnm{Maksim} \snm{Greiner}}
\author[
addressref={lmu,ec},                   
email={Frederik.Beaujean@lmu.de}   
]{\inits{FB}\fnm{Frederik} \snm{Beaujean}}
\author[
addressref={maxplanck,lmu},                   
email={ensslin@mpa-garching.mpg.de}   
]{\inits{TE}\fnm{Torsten} \snm{Enßlin}}


\address[id=maxplanck]{
  \orgname{Max Planck Institut für Astrophysik}, 
  \street{Karl-Schwarzschild-Strasse 1},                     %
  \postcode{85741}                                
  \city{Garching},                              
  \cny{Germany}                                    
}

\address[id=lmu]{
    \orgname{Ludwig-Maximilians-Universität München}, 
    \street{Geschwister-Scholl-Platz 1},                     %
    \postcode{80539}                                
    \city{München},                              
    \cny{Germany}                                    
}
\address[id=ec]{
    \orgname{Exzellenzcluster Universe}, 
    \street{Boltzmannstrasse 2},                     %
    \postcode{85748}                                
    \city{Garching},                              
    \cny{Germany}                                    
}

\begin{artnotes}
\end{artnotes}

\end{fmbox}


\begin{abstractbox}

\begin{abstract} 
    We introduce \ddo, a Python module for cluster-distributed multi-dimensional numerical arrays.
    It acts as a layer of abstraction between the algorithm code and the data-distribution logic.
    The main goal is to achieve usability without losing numerical performance and scalability.
    \ddo's global interface is similar to the one of a \verb|numpy.ndarray|, whereas the cluster node's local data is directly accessible for use in customized high-performance modules.
    \ddo\ is written in pure Python which makes it portable and easy to use and modify.
    Expensive operations are carried out by dedicated external libraries like \emph{numpy} and \emph{mpi4py}.
    The performance of \ddo\ is on a par with numpy for serial applications and scales well when moving to an MPI cluster.
    \ddo\ is open-source software available under the GNU General Public License v3 (GPL-3) at \url{https://gitlab.mpcdf.mpg.de/ift/D2O}.
%
\end{abstract}




\end{abstractbox}
%

\end{frontmatter}



\section{Introduction}
\subsection{Background}\label{sec:Backround}
Data sets in simulation and signal-reconstruction applications easily reach sizes too large for a single computer's random access memory (RAM).
A reasonable grid size for such tasks like
    galactic density reconstructions~\cite{greiner.electron_density} or
    multi-frequency imaging in radio astronomy~\cite{junklewitz.resolve}
    is a cube with a side resolution of $2048$.
Such a cube contains $2048^3 \approx 8.6 \cdot 10^9$ voxels.
Storing a 64-bit double for every voxel therefore consumes $64~\mathrm{GiB}$.
In practice one has to handle several or even many instances of those arrays which ultimately prohibits the use of single shared memory machines.
Apart from merely holding the arrays' data in memory,
    parallelization is needed to process those huge arrays within reasonable time.
This applies to basic arithmetics like addition and multiplication as well as to complex operations like Fourier transformation and advanced linear algebra,
    e.g. operator inversions or singular value decompositions.
Thus parallelization is highly advisable for code projects that must be scaled to high resolutions.

To be specific, the initial purpose of \ddo\ was to provide parallelization to the package for Numerical Information Field Theory (\nifty)\cite{nifty},
    which permits the abstract and efficient implementation of sophisticated signal processing methods.
Typically, those methods are so complex on their own that a \nifty\ user should not need to bother with parallelization details in addition to that.
It turned out that providing a generic encapsulation for parallelization to \nifty\ is not straightforward
    as the applications \nifty\ is used for are highly diversified.
The challenge hereby is that, despite all their peculiarities, for those applications numerical efficiency is absolutely crucial.
Hence, for encapsulating the parallelization effort in \nifty\ we needed an approach that is flexible enough to adapt to those different applications such that numerical efficiency can be preserved: \ddo.

\ddo\ is implemented in Python.
As a high-level language with a low-entry barrier Python is widely used in computational science.
It has a huge standard library and an active community for 3rd party packages.
For computationally demanding applications Python is predominantly used as a steering language for external compiled modules because Python itself is slow for numerics.

This article is structured as follows.
Section \ref{sec:aim} gives the aims of \ddo,
    and section \ref{sec:alternatives} describes alternative data distribution packages.
We dicuss the code architecture in section \ref{sec:code_arch} ,
    the basic usage of \ddo\ in section \ref{sec:basic_usage},
    and the numerical scaling behavior in section \ref{sec:performance_scalability}.
Section \ref{sec:summary} contains our conclusion
    and appendix \ref{sec:adv_usage_func_behav} describes the detailed usage of \ddo.

\subsection{Aim}\label{sec:aim}
As most scientists are not fully skilled software engineers, for them the hurdle for developing parallelized code is high.
Our goal is to provide data scientists with a numpy array-like object (cf.\ \emph{numpy}~\cite{walt.numpy}) that distributes data among several nodes of a cluster in a user-controllable way.
The user, however, shall not need to have profound knowledge about parallel programming with a system like \emph{MPI}~\cite{mpi-1-standard, mpi-2-standard} to achieve this.
The transition to use \distributeddataobjects\ instead of numpy arrays in existing code must be as straightforward as possible.
Hence, \ddo\ shall in principle run -- at least in a non-parallelized manner -- with standard-library dependencies available;
    the packages needed for parallel usage should be easily available.
Whilst providing a global-minded interface, the node's local data should be directly accessible in order to enable the usage in specialized high-performance modules.
This approach matches with the theme of \emph{\mbox{DistArray}} \cite{distarray}: ``Think globally, act locally''.
Regarding \ddo's architecture we do not want to make any a-priori assumptions about the specific distribution strategy, but retain flexibility:
    it shall be possible to adapt to specific constraints induced from third-party libraries a user may incorporate.
For example, a library for fast Fourier transformations like \emph{FFTW}~\cite{Frigo:1999:FFT:301618.301661} may rely on a different data-distribution model than
    a package for linear algebra operations like \emph{ScaLAPACK}~\cite{slug}\footnote{FFTW distributes slices of data, while ScaLAPACK uses a block-cyclic distribution pattern.}.
In the same manner it shall not matter whether a new distribution scheme stores data redundantly or not, e.g. when a node is storing not only a distinct piece of a global array, but also its neighboring (ghost) cells \cite{DadoneGhostCellGrid}.

Our main focus is on rendering extremely costly computations possible in the first place;
    not on improving the speed of simple computations that can be done serially.
Although primarily geared towards \emph{weak scaling}, it turns out that \ddo{} performs very well in \emph{strong-scaling} scenarios, too; see section \ref{sec:performance_scalability} for details.

\subsection{Alternative Packages}\label{sec:alternatives}
There are several alternatives to \ddo. We discuss the differences to \ddo\ and why the alternatives are not sufficient for our needs.

\subsubsection{DistArray}
\emph{DistArray}~\cite{distarray} is very mature and powerful.
Its approach is very similar to \ddo:
    It mimics the interface of a multi dimensional numpy array while distributing the data among nodes in a cluster.
However, DistArray involves a design decision that makes it inapt for our purposes:
    it has a strict client-worker architecture.
DistArray either needs an \emph{ipython ipcluster}~\cite{PER-GRA:2007.ipython} as back end or must be run with two or more MPI processes.
The former must be started before an interactive ipython session is launched.
This at least complicates the workflow in the prototyping phase
    and at most is not practical for batch system based computing on a cluster.
The latter enforces tool-developers who build on top of \emph{DistArray} to demand that their code always is run parallelized.
Both scenarios conflict with our goal of minimal second order dependencies and maximal flexibility,
    cf.\ section \ref{sec:aim}.
Nevertheless, its theme also applies to \ddo: ``Think globally, act locally''.

\subsubsection{scalapy (ScaLAPACK)}
\emph{scalapy} is a Python wrapper around \emph{ScaLAPACK}~\cite{slug}, which is ``a library of high-performance linear algebra routines for parallel distributed memory machines''~\cite{scalapack.online}.
The \verb|scalapy.DistributedMatrix| class essentially uses the routines from ScaLAPACK and therefore is limited to the functionality of that:
    two-dimensional arrays and very specific block-cyclic distribution strategies that optimize numerical efficiency in the context of linear algebra problems.
In contrast, we are interested in $n$-dimensional arrays whose distribution scheme shall be arbitrary in the first place.
Therefore scalapy is not extensive enough for us.

\subsubsection{petsc4py (PETSc)}
\emph{petsc4py} is a Python wrapper around \emph{PETSc}, which ``is a suite of data structures and routines for the scalable (parallel) solution of scientific applications modeled by partial differential equations''~\cite{petsc-web-page}.
Regarding distributed arrays its scope is as focused as scalapy to its certain problem domain --
    here: solving partial differential equations.
The class for distributed arrays \verb|petsc4py.PETSc.DMDA| is limited to one, two and three dimensions
    as PETSc uses a highly problem-fitted distribution scheme.
We in contrast need $n$-dimensional arrays with arbitrary distribution schemes.
Hence, petsc4py is not suitable for us.

\section{Code Architecture}\label{sec:code_arch}
\subsection{Choosing the Right Level of Parallelization}
\ddo\ distributes numerical arrays over a cluster in order to parallelize and therefore to speed up operations on the arrays themselves.
An application that is built on top of \ddo\ can profit from its fast array operations that may be performed on a cluster.
However, there are various approaches how to deploy an algorithm on a cluster and
    \ddo\ implements only one of them.
In order to understand the design decisions of \ddo\
    and its position respective to other packages, cf.\ section \ref{sec:alternatives},
    we will now discuss the general problem setting of parallelization and possible approaches for that.
Thereby we reenact the decision process which led to the characteristics \ddo\ has today.\\

\subsubsection{Vertical \& Horizontal Scaling}
Suppose we want to solve an expensive numerical problem which involves operations on data arrays.
To reduce the computation time one can in principle do two things.
Either use a faster machine -- \emph{vertical scaling} --
    or use more than one machine -- \emph{horizontal scaling}.
Vertical scaling has the advantage that existing code does not need to be changed\footnote{This is true if scaling up does not involve a change of the processor architecture.},
    but in many cases this is not appropriate.
Maybe one already uses the fastest possible machine,
    scaling up is not affordable
    or even the fastest machine available is still too slow.
Because of this, we choose horizontal scaling.
\subsubsection{High- \& Low-Level Parallelization}
With horizontal scaling we again face two choices:
    high- and low-level parallelization.
With high-level parallelization, many replicas of the algorithm run simultaneously,
    potentially on multiple machines.
Each instance then works independently if possible,
    solving an isolated part of the global problem.
At the end, the individual results get collected and merged.
The python framework \emph{pathos}~\cite{mckerns.pathos} provides functionality for this kind of procedure.\\
An example of high-level parallelization is a sample generator which draws from a probability distribution.
Using high-level parallelization many instances of the generator produce their own samples,
    which involves very little communication overhead.
The sample production process itself, however, is not sped up.\\
In low-level parallelization, several nodes work together on one basic task at a time.
For the above sample generator,
    this means that all nodes work on the same sample at a time.
Hence, the time needed for producing individual samples is reduced;
    they are serially generated by the cluster as a whole.
\subsubsection{Downsides}
Both of these approaches have their drawbacks.
For high-level parallelization the algorithm itself must be parallelizable.
Every finite algorithm has a maximum degree of intrinsic parallelization\footnote{For the exemplary sample generator the maximum degree of parallelization is the total number of requested samples.}.
If this degree is lower than the desired number of processes then high-level parallelization reaches its limits.
This is particularly true for algorithms that cannot be parallelized by themselves, like iterative schemes.
Furthermore, there can be an additional complication:
    if the numerical problem deals with extremely large objects it may be the case that it is not at all solvable by one machine alone\footnote{In case of the sample generator this would be the case if even one sample would be too large for an individual machine's RAM.}.\\
Now let us consider low-level parallelization.
As stated above,
    we assume that the solution of the given numerical problem involves operations on data arrays.
Examples for those are unary\footnote{E.g. the positive, negative or absolute values of the array's individual elements or the maximum, minimum, median or mean of all its elements.},
    binary\footnote{E.g. the sum, difference or product of two data arrays.} or
    sorting operations,
    but also more advanced procedures like Fourier transformations or (other)
    linear algebra operations.
Theoretically, the absolute maximum degree of intrinsic parallelization for an array operation is equal to the array's number of elements.
For comparison, the problems we want to tackle involve at least $10^8$ elements
    but most of the TOP500~\cite{TOP500} supercomputers possess $10^6$ cores or less.
At first glance this seems promising.
But with an increasing number of nodes that participate in one operation the computational efficiency may decrease considerably.
This happens if the cost of the actual numerical operations becomes comparable to the generic program and inter-node communication overhead.
The ratios highly depend on the specific cluster hardware and the array operations performed.

\subsubsection{Problem Sizes}
Due to our background in signal reconstruction and grid-based simulations, we decide to use low-level parallelization for the following reasons.
First, we have to speed up problems that one cannot parallelize algorithmically,
    like fixed-point iterations or step-wise simulations.
Second, we want to scale our algorithms to higher resolutions
    while keeping the computing time at least constant.
Thereby the involved data arrays become so big that a single computer would be oversubscribed.
Because of this, the ratio of \emph{array size} to \emph{desired degree of parallelization} does not become such that the computational efficiency would decrease considerably.
In practice we experience a good scaling behavior with up to $\approx 10^3$ processes\footnote{This was the maximum number of processes available for testing.} for problems of size $8192^2$,
    cf.\ section \ref{sec:performance_scalability}.
Hence, for our applications the advantages of low-level parallelization clearly outweigh its drawbacks.


\subsection{\ddo\ as Layer of Abstraction}
Compared to high-level parallelization,
    the low-level approach is more complicated to implement.
In the best case, for the former one simply runs the serial code in parallel on the individual machines;
    when finished one collects and combines the results.
For the latter, when doing the explicit coding one deals with local data portions of the global data array on the individual nodes of the cluster.
Hence, one has to keep track of additional information:
    for example, given a distribution scheme, which portion of the global data is stored on which node of the cluster?
Keeping the number of cluster nodes, the size and the dimensionality of the data arrays arbitrary implies a considerable complication for indexing purposes.
By this, while implementing an application one has to take care of two non-trivial tasks.
On the one hand, one must program the logic of distributing and collecting the data; i.e.\ the data handling.
On the other hand, one must implement the application's actual (abstract) algorithm.
Those two tasks are conceptually completely different and therefore a mixture of implementations should be avoided.
Otherwise there is the risk that features of an initial implementation -- like the data distribution scheme -- become hard-wired to the algorithm, inhibiting its further evolution.
Thus it makes sense to insert a layer of abstraction between the algorithm code and the data distribution logic.
Then the abstract algorithm can be written in a serial style from which all knowledge and methodology regarding the data distribution is encapsulated.
This layer of abstraction is \ddo.

\subsection{Choosing a Parallelization Environment}
To make the application spectrum of \ddo\ as wide as possible we want to maximize its portability and reduce its dependencies.
This implies that -- despite its parallel architecture -- \ddo\ must just as well run within a single-process environment for cases when no elaborate parallelization back end is available.
But nevertheless, \ddo\ must be massively scalable.
This relates to the question of which distribution environment should be used.
There are several alternatives:
\begin{itemize}
	\item Threading and multiprocessing:
        These two options limit the application to a single machine which conflicts with the aim of massive scalability.
	\item (py)Spark~\cite{Zaharia:2010:SCC:1863103.1863113} and hadoop~\cite{hadoop}: These modern frameworks are very powerful but regrettably too abstract for our purposes,
        as they prescind the location of individual portions of the full data.
    Building a numpy-like interface would be disproportionately hard or even unfeasible.
    In addition to that, implementing a low-level interface for highly optimized applications which interact with the node's local data  is not convenient within pySpark.
    Lastly, those frameworks are usually not installed as standard dependencies on scientific HPC clusters.
	\item MPI~\cite{mpi-1-standard, mpi-2-standard}:
    The \emph{Message Passing Interface} is available on virtually every HPC cluster via well-tested implementations like \emph{OpenMPI}~\cite{gabriel04:_open_mpi}, \emph{MPICH2}~\cite{mpich2} or \emph{\mbox{Intel MPI}}~\cite{intelmpi}.
    The open implementations are also available on commodity multicore hardware like desktops or laptops.
    A Python interface to MPI is given by the Python module \emph{mpi4py}~\cite{Dalcin20051108.mpi4py}.
    MPI furthermore offers the right level of abstraction for hands-on control of distribution strategies for the package developers.
\end{itemize}
Given these features we decide to use \emph{MPI} as the parallelization environment for \ddo.
We stress that in order to fully utilize \ddo{} on multiple cores, a user does not need to know how to program in MPI; it is only necessary to execute the program via MPI as shown in the example in section~\ref{sec:distr_arrays}.

\subsection{Internal Structure}\label{sec:internal_structure}
\subsubsection{Composed Object}\label{sec:composed_object}
A main goal for the design of \ddo\ was to make no a-priori assumptions about the specific distribution strategies that will be used in order to spread array data across the nodes of a cluster.
Because of this, \ddo's distributed array -- \verb|d2o.distributed_data_object| -- is a composed object; cf.\ figure \ref{fig_object_structure}.\\
The \distributeddataobject\ itself provides a rich user interface,
    and makes sanity and consistency checks regarding the user input.
In addition to that, the \distributeddataobject\ possesses an attribute called \verb|data|.
Here the MPI processes' local portion of the global array data is stored,
    even though the \distributeddataobject\ itself will never make any assumptions about its specific content
    since the distribution strategy is arbitrary in the first place.
The \distributeddataobject{} is the only object of the \ddo{} library that a casual user would interact with.\\
For all tasks that require knowledge about the certain distribution strategy every \distributeddataobject\ possesses an instance of a \verb|d2o.distributor| subclass.
This object stores all the distribution-scheme and cluster related information
    it needs in order to scatter (gather) data to (from) the nodes and to serve for special methods,
    e.g.\ the array-cumulative sum.
The \distributeddataobject{} builds its rich user interface on top of those abstracted methods of its distributor.\\
The benefit of this strict separation is that the user interface becomes fully detached from the distribution strategy;
    may it be block-cyclic or slicing,
    or have neighbor ghost cells or not, et cetera.
Currently there are two fundamental distributors available:
    a generic \emph{slicing}-\footnote{The slicing is done along the first array axis.} and a \emph{not}-distributor.
From the former, three special slicing distributors are derived:
    \emph{fftw}\footnote{The fftw-distributor uses routines from the pyFFTW~\cite{pyfftw, Frigo:1999:FFT:301618.301661} package~\cite{Frigo:1999:FFT:301618.301661} for the data partitioning.},
    \emph{equal}\footnote{The equal-distributor tries to split the data in preferably equal-sized parts.} and
    \emph{freeform}\footnote{The local data array's first axis is of arbitrary length for the freeform-distributor.}.
The latter, the \emph{not}-distributor, does not do any data-distribution or -collection but stores the full data on every node redundantly.

\subsubsection{Advantages of a Global View Interface}
\ddo's global view interface makes it possible to build software that remains completely independent from the distribution strategy and the used number of cluster processes.
This in turn enables the development of 3rd party libraries that are very end-use-case independent.
An example for this may be a mathematical optimizer;
    an object which tries to find for a given scalar function $f$ an input vector $\vec{x}$ such that the output $y = f(\vec{x})$ becomes minimal.
It is interesting to note that many optimization algorithms solely use basic arithmetics like vector addition or scalar multiplication when acting on $\vec{x}$.
As such operations act locally on the elements of an array,
    there is no preference for one distribution scheme over another
    when distributing $\vec{x}$ among nodes in a cluster.
Two different distribution schemes will yield the same performance
    if their load-balancing is on a par with each other.
Further assume that $f$ is built on \ddo, too.
On this basis, one could now build an application that uses the minimizer
    but indeed has a preference for a certain distribution scheme.
This may be the case if the load-balancing of the used operations is non-trivial
    and therefore only a certain distribution scheme guarantees high evaluation speeds.
While the application's developer therefore enforces this scheme,
    the minimizer remains completely unaffected by this as it is agnostic of the array's distribution strategy.

\section{Basic Usage}\label{sec:basic_usage}
In the subsequent sections we will illustrate the basic usage of \ddo{} in order to explain its functionality and behavior.
A more extended discussion is given in Appendix \ref{sec:adv_usage_func_behav}.
Our naming conventions are:
\begin{itemize}
	\item instances of the \verb|numpy.ndarray| class are labeled \verb|a| and \verb|b|,
	\item instances of \verb|d2o.distributed_data_object| are labeled \verb|obj| and \verb|p|.
\end{itemize}
In addition to these examples, the interested reader is encouraged to have a look into the \distributeddataobject{} method's docstrings for further information; cf. the project's web page \url{https://gitlab.mpcdf.mpg.de/ift/D2O}.\\

\subsection{Initialization}
Here we discuss how to initialize a \distributeddataobject{} and compare some of its basic functionality to that of a \verb|numpy.ndarray|.
First we import the packages.
\begin{lstlisting}[language=iPython]
	|\iin| import numpy as np
	|\iin| from d2o import distributed_data_object
\end{lstlisting}
Now we set up some test data using numpy.
\begin{lstlisting}[language=iPython]
    |\iin| a = np.arange(12).reshape((3, 4))
    |\iin| a
    |\out| array([[ 0,  1,  2,  3],
                  [ 4,  5,  6,  7],
                  [ 8,  9, 10, 11]])
\end{lstlisting}
One way to initialize a \distributeddataobject{} is to pass an existing numpy array.
\begin{lstlisting}[language=iPython]
    |\iin| obj = distributed_data_object(a)
    |\iin| obj
    |\out| <distributed_data_object>
           array([[ 0,  1,  2,  3],
                   [ 4,  5,  6,  7],
                   [ 8,  9, 10, 11]])
\end{lstlisting}
The output of the \verb|obj| call shows the local portion of the global data available in this process.

\subsection{Arithmetics}
Simple arithmetics and point-wise comparison work as expected from a numpy array.
\begin{lstlisting}[language=iPython]
    |\iin| (2*obj, obj**3, obj >= 5)
    |\out| (<distributed_data_object>
            array([[ 0,  2,  4,  6],
                   [ 8, 10, 12, 14],
                   [16, 18, 20, 22]]),
            <distributed_data_object>
            array([[   0,    1,    8,   27],
                   [  64,  125,  216,  343],
                   [ 512,  729, 1000, 1331]]),
            <distributed_data_object>
            array([[False, False, False, False],
                   [False,  True,  True,  True],
                   [ True,  True,  True,  True]], dtype=bool))
\end{lstlisting}
Please note that the \distributeddataobject{} tries to avoid inter-process communication whenever possible.
Therefore the returned objects of those arithmetic operations are instances of \distributeddataobject{}, too.
However, the \ddo{} user must be careful when combining \distributeddataobjects{} with numpy arrays.
If one combines two objects with a binary operator in Python (like \verb|+|, \verb|-|, \verb|*|, \verb|\|, \verb|%| or \verb|**|)
    it will try to call the respective method (\verb|__add__|, \verb| __sub__|, etc...) of the \emph{first} object.
If this fails, i.e. if it throws an exception,
    Python will try to call the \emph{reverse} methods of the \emph{second} object (\verb|__radd__|, \verb|__rsub__|, etc...):
\begin{lstlisting}[language=iPython]
    |\iin| a + 1; # calls a.__add__(1) -> returns a numpy array
    |\iin| 1 + a; # 1.__add__ not existing -> a.__radd__(1)
\end{lstlisting}
Depending on the conjunction's ordering, the return type may vary when combining numpy arrays with \distributeddataobjects{}.
If the numpy array is in the first place,
    numpy will try to extract the second object's array data using its \verb|__array__| method.
This invokes the \distributeddataobject's \verb|get_full_data| method that communicates the full data to every process.
For large arrays this is extremely inefficient and should be avoided by all means.
Hence, it is crucial for performance to assure that the \distributeddataobject's methods will be called by Python.
In this case, the locally relevant parts of the array are extracted from the numpy array and then efficiently processed as a whole.
\begin{lstlisting}[language=iPython]
    |\iin| a + obj # numpy converts obj -> inefficient
    |\out| array([[ 0,  2,  4,  6], # note: numpy.ndarray
                  [ 8, 10, 12, 14],
                  [16, 18, 20, 22]])

    |\iin| obj + a # obj processes a -> efficient
    |\out| <distributed_data_object>
            array([[ 0,  2,  4,  6],
                   [ 8, 10, 12, 14],
                   [16, 18, 20, 22]])
\end{lstlisting}
\subsection{Array Indexing}
The \distributeddataobject{} supports most of numpy's indexing functionality,
    so it is possible to work with scalars, tuples, lists, numpy arrays and \distributeddataobjects{} as input data.
Every process will extract its locally relevant part of the given data-object and then store it;
    cf.\ \ref{sec:advanced_indexing}.
\begin{lstlisting}[language=iPython]
    |\iin| obj
    |\out| <distributed_data_object>
            array([[ 0,  1,  2,  3],
                   [ 4,  5,  6,  7],
                   [ 8,  9, 10, 11]])

    |\iin| obj[1] # extract a row
    |\out| <distributed_data_object>
            array([4, 5, 6, 7])

    |\iin| obj[1,-2] # extract single entry
    |\out| 6

    |\iin| obj[::2, 1::2] # slicing notation
    |\out| <distributed_data_object>
            array([[ 1,  3],
                   [ 9, 11]])

    # sets data using slicing
    |\iin| obj[::2, 1::2] = [[111, 222], [333, 444]]
    |\iin| obj
    |\out| <distributed_data_object>
            array([[  0, 111,   2, 222],
                   [  4,   5,   6,   7],
                   [  8, 333,  10, 444]])
\end{lstlisting}
By default it is assumed that all processes use the \emph{same} key-object when accessing data.
See section \ref{sec:local_keys} for more details regarding process-individual indexing.

\subsection{Distribution Strategies}
In order to specify the distribution strategy explicitly one may use the ``\verb|distribution_strategy|'' keyword:
\begin{lstlisting}[language=iPython]
    |\iin| obj = distributed_data_object(
                    a, distribution_strategy='equal')
    |\iin| obj.distribution_strategy
    |\out| 'equal'
\end{lstlisting}
See section \ref{sec:distribution_strategies} for more information on distribution strategies.

\subsection{Distributed Arrays} \label{sec:distr_arrays}
To use \ddo{} in a distributed manner, one has to create an MPI job.
This example shows how four MPI processes hold individual parts of the global data and how distributed read \& write access works.
The script is started via the command:
\begin{verbatim}
mpirun -n 4 python get_set_data.py
\end{verbatim}

\begin{lstlisting}[language=iPython]
    # get_set_data.py
    from mpi4py import MPI
    import numpy as np
    from d2o import distributed_data_object
    # Get the process' rank number (0,1,2,3) from MPI
    rank = MPI.COMM_WORLD.rank

    # Initialize some data
    a = np.arange(16).reshape((4,4))
    # Initialize the distributed_data_object
    obj = distributed_data_object(a) |\label{listing_dist_arrays_1}|

    # Print the process' local data
    print (rank, obj.get_local_data()) |\label{listing_dist_arrays_2}|
    # extract data via slicing
    print (rank, obj[0:3:2, 1:3].get_local_data()) |\label{listing_dist_arrays_3}|

    b = -np.arange(4).reshape((2,2))
    obj[2:4,1:3] = b # Write b into obj |\label{listing_dist_arrays_4}|

    # Print the process' local data
    print (rank, obj.get_local_data())

    # Consolidate the data
    full_data = obj.get_full_data() |\label{listing_dist_arrays_5}|
    if rank == 0: print (rank, full_data)
\end{lstlisting}
The \distributeddataobject{} gets initialized in line \ref{listing_dist_arrays_1} with the following array:
\begin{verbatim}
array([[ 0,  1,  2,  3],
       [ 4,  5,  6,  7],
       [ 8,  9, 10, 11],
       [12, 13, 14, 15]]))
\end{verbatim}
Here, the script is run in four MPI processes; cf.\ \verb|mpirun -n 4 [...]|.
The data is split along the first axis; the print statement in line \ref{listing_dist_arrays_2} yields the four pieces:
\begin{verbatim}
(0, array([[ 0,  1,  2,  3]]))
(1, array([[ 4,  5,  6,  7]]))
(2, array([[ 8,  9, 10, 11]]))
(3, array([[12, 13, 14, 15]]))
\end{verbatim}
The second print statement (line \ref{listing_dist_arrays_3}) illustrates the behavior of data extraction;
    \verb|obj[0:3:2, 1:3]| is slicing notation for the entries 1, 2, 9 and 10\footnote{This notation can be decoded as follows. The numbers in a slice correspond to \verb|start:stop:step| with \verb|stop| being exclusive. \verb|obj[0:3:2, 1:3]| means to take every second line from the lines 0, 1 and 2, and to then take from this the elements in collumns 1 and 2.}.
This expression returns a \distributeddataobject{} where the processes possess the individual portion of the requested data.
This means that the distribution strategy of the new (sub-)array is determined by and aligned to that of the original array.
\begin{verbatim}
(0, array([[1, 2]]))
(1, array([], shape=(0, 2), dtype=int64)) # empty
(2, array([[ 9, 10]]))
(3, array([], shape=(0, 2), dtype=int64)) # empty
\end{verbatim}
The result is a \distributeddataobject\ where the processes 1 and 3 do not possess any data as they had no data to contribute to the slice in \verb|obj[0:3:2, 1:3]|.
In line \ref{listing_dist_arrays_4} we store a small 2x2 block \verb|b| in the lower middle of \verb|obj|. The process' local data reads:
\begin{verbatim}
(0, array([[ 0,  1,  2,  3]]))
(1, array([[ 4,  5,  6,  7]]))
(2, array([[ 8,  0, -1, 11]]))
(3, array([[12, -2, -3, 15]]))
\end{verbatim}
Finally, in line \ref{listing_dist_arrays_5} we use \verb|obj.get_full_data()| in order to consolidate the distributed data;
    i.e. to communicate the individual pieces between the processes and merge them into a single numpy array.
\begin{verbatim}
(0, array([[ 0,  1,  2,  3],
           [ 4,  5,  6,  7],
           [ 8,  0, -1, 11],
           [12, -2, -3, 15]]))
\end{verbatim}

\section{Performance and Scalability}\label{sec:performance_scalability}
In this section we examine the scaling behavior of a \distributeddataobject\ that uses the \emph{equal} distribution strategy.
The timing measurements were performed on the C2PAP Computing Cluster~\cite{c2pap}\footnote{The C2PAP computing cluster consists of 128 nodes, each possessing two Intel Xeon CPU E5-2680 (8 cores \@ 2.70~GHz + hyper-threading, 64~KiB L1 per core, 256~KiB L2 cache per core, 20~MiB L3 cache shared for all 8 cores) and 64 GiB RAM each. The nodes are connected via Mellanox Infiniband 40 Gbits/s}.
The software stack was built upon \emph{Intel MPI 5.1}, \emph{mpi4py 2.0}, \emph{numpy 1.11} and \emph{python 2.7.11}.
For measuring the individual timings we used the Python standard library module \emph{timeit} with a fixed number of $100$ repetitions.

Please note that \ddo\ comes with an extensive suite of unit tests featuring a high code-coverage rate.
By this we assure \ddo's continuous code development to be very robust
    and provide the users with a reliable interface definition.

Regarding \ddo's performance there are two important scaling directions:
    the size of the array data and
    the number of MPI processes.
One may distinguish three different contributions to the overall computational cost.
First, there is data management effort that every process has to cover itself.
Second, there are the costs for inter-MPI-process communication.
And third, there are the actual numerical costs.

\ddo\ has size-independent management overhead compared to numpy.
Hence, the larger the arrays are for which \ddo\ is used, the more efficient the computations become.
We will see below that there is a certain array size per node
    -- roughly $2^{16}$ elements --
    from which on d2o's management overhead becomes negligible compared to the purely numerical costs.
This size corresponds to a two-dimensional grid with a resolution of $256 \times 256$ or equivalently $0.5~\mathrm{MiB}$ of 64-bit doubles.
In section \ref{sec:scale_size} we focus on this very ratio of management overhead to numerical costs.

\ddo\ raises the claim to be able to operate well running with a single process as well as in a highly parallelized regime.
In section \ref{sec:scale_nodes}, the scaling analysis regarding the MPI process count is done with a fixed local array size for which the process overhead is negligible compared to the numerical costs.
For this weak scaling analysis we are interested in the costs arising from inter-process communication compared to those of actual numerics.

In the following three sections, we study the strong scaling of \ddo{},
    where the performance is a result of the combination of all three cost contributions.
Section \ref{sec:strong_scaling} covers the case in which the number of MPI processes is increased while the array size is left constant.
In section \ref{sec:comp_distarray} we compare \ddo's to \emph{DistArray}'s\cite{distarray} performance
and finally, in section \ref{sec:wiener} we benchmark \ddo's strong-scaling behavior when applied to a real-world application:
    a Wiener filter signal reconstruction.

A discussion on \ddo's efficient Python iterators can be found in appendix \ref{sec:perf_iter}.

\subsection{Scaling the Array Size}\label{sec:scale_size}
One may think of \ddo\ as a layer of abstraction that is added to numpy arrays
    in order to take care of data distribution and collection among multiple MPI processes.
This abstraction comes with inherent Python overhead, separately for each MPI process.
Therefore, if one wants to analyze how the ratio of \emph{management overhead} to \emph{actual numerical effort} varies with the data size,
    only the individual process' data size is important.
Because of this, all timing tests for this subsection were carried out with one MPI process only.

A common task during almost all numerical operations is to create a new array object for storing its results\footnote{Exceptions to this are inplace operations which reuse the input array for the output data.}.
Hence, the speed of object creation can be crucial for overall performance there.
Note that in contrast to a numpy array which basically just allocates RAM,
    several things must be done during the initialization of a \distributeddataobject.
The Python object instance itself must be created,
    a distributor must be initialized which involves parsing of user input,
    RAM must be allocated,
    the \distributeddataobject\ must be registered with the d2o\_librarian (cf.\ \ref{sec:librarian}),
    and, if activated, inter-MPI-process communication must be done for performing sanity checks on the user input.

By default the initialization requires \mus{60} to complete for a \distributeddataobject\ with a shape of \verb|(1,)| when run within one single MPI process.
Using this trivial shape makes the costs for memory allocation negligible compared to the others tasks.
Hence, those \mus{60} represent \ddo's constant overhead compared to numpy,
    since a comparable numpy array requires $\approx \mus{0.4}$ for initialization.

In order to speed up the initialization process one may disable all sanity checks on the user input that require MPI communication,
    e.g. if the same datatype was specified in all MPI processes.
Even when run with one single process,
    skipping those checks reduces the costs by \mus{27} from \mus{60} to \mus{33}.

Because of the high costs, it is important to avoid building \distributeddataobjects\ from scratch over and over again.
A simple measure against this is to use inplace operations like \verb|obj += 1| instead of \verb|obj = obj + 1| whenever possible.
This is generally a favorable thing to do -- also for numpy arrays --
    as this saves the costs for repeated memory allocation.
Nonetheless, also non-inplace operations can be improved in many cases,
    as often the produced and the initial \distributeddataobject\ have all of their attributes in common, except for their data:
    they are of the same shape and datatype, and use the same distribution strategy and MPI communicator;
    cf.\ \verb|p = obj + 1|.
With \verb|obj.copy()| and \verb|obj.copy_empty()| there exist two cloning methods that we implemented to be as fast as allowed by pure Python.
Those methods reuse as much already initialized components as possible and are therefore faster than a fresh initialization:
    for the \distributeddataobject\ from above
    \verb|obj.copy()| and \verb|obj.copy_empty()| consume \mus{7.9} and \mus{4.3}, respectively.

Table \ref{table_scaling_size} shows the performance ratio in percent between serial \ddo\ and numpy.
The array sizes range from $2^0=1$ to $2^{25} \approx 3.3\cdot10^7$ elements.
In the table, $100\%$ would mean that \ddo\ is as fast as numpy.

The previous section already suggested that for tasks that primarily consist of initialization work -- like \emph{array creation} or \verb|copy_empty| -- \ddo\ will clearly follow behind numpy.
However, increasing the array size from $2^{20}$ to $2^{22}$ elements implies a considerable performance drop for numpy's memory allocation.
This in turn means that for arrays with more than $2^{22}$ elements \ddo's relative overhead becomes less significant: e.g.\ \verb|np.copy_empty| is then only a factor of four faster than \verb|obj.copy_empty()|.

Functions like \verb|max| and \verb|sum| return a scalar number;
    no expensive return-array must be created.
Hence, \ddo's overhead is quite modest:
    even for size $1$ arrays, \ddo's relative performance lies above $50\%$.
Once the size is greater than $2^{18}$ elements the performance is higher than $95\%$.

\verb|obj[::-2]| is \emph{slicing syntax} for \emph{``take every second element from the array in reverse order''}.
It illustrates the costs of the data-distribution and collection logic
    that even plays a significant role if there is no inter-process communication involved.
Again, with a large-enough array size, \ddo's efficiency becomes comparable to that of numpy.

Similarly to \verb|obj[::-2]|, the remaining functions in the table return a \distributeddataobject{} as their result and therefore suffer from its initialization costs.
However, with an array size of $2^{16}$ elements and larger \ddo's relative performance is at least greater than approximately $65\%$.

An interesting phenomenon can be observed for \verb|obj + 0| and \verb|obj + obj|:
As for the other functions, their relative performance starts to increase significantly when an array size of $2^{16}$ is reached.
However, in contrast to \verb|obj += obj| which then immediately scales up above $95\%$,
    the relative performance of the non-inplace additions temporarily \emph{decreases} with increasing array size.
This may be due to the fact that given our test scenario $2^{18}$ elements almost take up half of the cache of C2PAP's Intel E5-2680 CPUs.
\ddo's memory overhead is now responsible for the fact, that its non-inplace operations -- which need twice the initial memory -- cannot profit that strongly from the cache anymore,
    whereas the numpy array still operates fast.
Once the array size is above $2^{22}$ elements numpy's just as \ddo's array-object is too large for profiting from the cache and therefore become comparably fast again:
    the relative performance is then greater than $98\%$.

Thus, when run within a single process, \ddo\ is ideally used for arrays larger than $2^{16}=65536$ elements which corresponds to $512~\mathrm{KiB}$.
From there the management overhead becomes less significant than the actual numerical costs.

\subsection{Weak Scaling: Proportional Number of Processes and Size of Data}\label{sec:scale_nodes}
Now we analyze the scaling behavior of \ddo\ when run with several MPI processes.
Repeating the beginning of section \ref{sec:performance_scalability}, there are three contributions to the execution time.
First, the fixed management overhead that every process has to cover itself, second, the communication overhead and third, the actual numerical costs.
In order to filter out the effect of a changing contribution of management overhead, in this section we fix the MPI processes' local array size to a fixed value.
Hence, now the global data size is proportional to the number of MPI processes.

Table \ref{table_scaling_nodes} shows the performance of various array operations normalized to the time \ddo\ needs when running with one process only.
Assuming that \ddo\ had no communication overhead and an operation scaled perfectly linearly with array size, the performance would be rated at $100\%$.

In theory, operations that do not inherently require inter-process communication like point-wise array addition or subtraction ideally scale linearly.
And in fact,
    \ddo\ scales excellently with the number of processes involved for those functions:
    here we tested \verb|copy|, \verb|copy_empty|, \verb|sum(axis=1)|, \verb|obj + 0|, \verb|obj + obj|, \verb|obj += obj| and \verb|sqrt|.

Comparing \verb|sum(axis=0)| with \verb|sum(axis=1)| illustrates the performance difference between those operations that involve inter-process communication and those that don't:
    the \emph{equal} distribution strategy slices the global array along its first axis in order to distribute the data among the individual MPI processes.
Hence, \verb|sum(axis=0)| -- which means to take the sum along the first axis -- does intrinsically involve inter-process communication
    whereas \verb|sum(axis=1)| does not.
Similarly to \verb|sum(axis=0)| also the remaining functions in table \ref{table_scaling_nodes} are affected by an increasing number of processes as they involve inter-process communication.

But still, even if -- for example in case of \verb|sum(axis=0)| -- the relative performance may drop to $28.2\%$ when using 256 processes,
    this means that the operation just took 3.5 times longer than the single-process run,
    whereat the array size has been increased by a factor of $256$.
This corresponds to a speedup factor of $72.2$.

\subsection{Strong Scaling: Varying Number of Processes with a Fixed Size of Data}\label{sec:strong_scaling}
Similarly to the previous section \ref{sec:scale_nodes}, we vary the number of processes but now fix the data size \emph{globally} instead of locally.
This corresponds to the real-life scenario in which the problem size and resolution are already set -- maybe by environmental conditions --
    and now one tries to reduce the run time by using more cores.
Since the size of the local data varies with the number of processes,
    the overall scaling behavior is now a mixture of the varying ratio between management overhead and process-individual numerical costs,
    and the fact that an increasing amount of CPU cache becomes available at the expense of increased communication effort.
Table \ref{table_scaling_nodes_strong} shows the benchmarking results for the same set of operations as used in the previous section on weak scaling
    and the results are reasonable.
Those operations in the list that inherently cannot scale strongly as they consist of purely node-individual work, namely the \verb|initialization| and \verb|copy_empty|, show that their performance just does not increase with the number of processes.
In contrast, operations without communication effort benefit strongly from the increasing total amount of CPU cache combined with smaller local arrays;
    above all \verb|copy| which is about 3 times faster than what one would expect from linear scaling  to 256 processes\footnote{This very strong scaling is indeed realistic: when analyzing pure numpy arrays one gets speedups in the order of magnitude of even $800~\%$.}.

In theory, the strong-scaling behavior is the combination of the size- and weak-scaling we discussed in sections \ref{sec:scale_size} and \ref{sec:scale_nodes}.
In order to verify whether the strong-scaling behavior makes sense, we estimate the strong-scaling performance using the information from size- and weak-scaling.

We choose the \verb|sum()| method as our test case.
During the reduction phase, the $n$ MPI-processes exchange their local results with each other.
This corresponds to adding $n$ times a fixed amount of communication time to the total computing time needed.
Hence, we perform a linear fit to the weak-scaling data; cf.\ table~\ref{table_scaling_nodes}.
Furthermore, we assume that the local computation time is roughly proportional to the local-array size.
This is true for sufficiently large array sizes,
    since then numpy scales linearly
    and \ddo\ is in a good efficiency regime, cf.\ \ref{table_scaling_size}.
Again we performed a linear fit but now on the size-scaling timing data.
Combining those two linear fits leads to the following run-time formula for applying \verb|sum()| to an array with shape $(4096, 4096)$:
\begin{equation}
t(n) = (0.0065n + 1.57/n)~\mathrm{s}
\end{equation}
In the case of linear scaling, $t(n)$ is expected to be equal to $t(1)/n$.
Hence, the relative performance $p(n)$ is the ratio between the two:
\begin{equation}
p(n)_{\mathrm{estimated}} = \frac{\nicefrac{t(1)}{n}}{t(n)} = \frac{241.8}{240.8 + n^2}
\end{equation}
Comparing the estimate with the actually measured relative performance -- cf.\ table \ref{table_scaling_nodes_strong} -- table \ref{tab_strong_scaling_estimate} shows that even under those rough assumptions the strong scaling behavior of \verb|sum()| can be explained as the combination of size- and weak scaling within about 20 \% accuracy.

\begin{table}[h!]
\caption{Strong Scaling Behavior: Estimate vs.\ Measurement}\label{tab_strong_scaling_estimate}
      \begin{tabular}{c|cccccccc}
            $\#\mathrm{processes:~} n$ & 1  & 4 & 8 & 16 & 32 & 64 & 128 & 256 \\ \hline
            $p(n)_{\mathrm{estimated}}$& 100 \% & 94.2 \% & 79.3 \% & 48.7 \% & 19.1 \% & 5.58 \% & 1.45 \% & 0.37 \% \\
            $p(n)_{\mathrm{measured}}$ & 100 \% & 91.7 \% & 74.1 \% & 60.9 \% & 24.4 \% & 7.05 \% & 1.82 \% & 0.45 \% \\
      \end{tabular}
\end{table}

\subsection{Strong Scaling: Comparison with DistArray}\label{sec:comp_distarray}
In section \ref{sec:alternatives} we discussed several competitors to \ddo{}.
Because of their similarities, we conducted the strong scaling tests -- as far as possible\footnote{DistArray does, for example, not support negative step-sizes for slicing (\texttt{[::-2]}) and also the special method \texttt{bincount} is not available.} -- also with DistArray and compare the performance.
In table \ref{table_scaling_nodes_distarray} the results are shown for the subset of all operations from the previous sections that where available for DistArray, too.

While being at least on a par for numerical operations when being run single-threaded,
    \ddo\ outperforms DistArray more and more with an increasing number of processes.
Furthermore, it is conspicuous that DistArray does not seem to support inplace array operations.
Because of this, the inplace addition \verb|obj+=obj| is way slower with DistArray than with \ddo\ which is on a par with numpy in most cases, cf. tables \ref{table_scaling_size}, \ref{table_scaling_nodes} and \ref{table_scaling_nodes_strong}.

The fact that \ddo\ is way more efficient when doing numerics on very small arrays -- like \verb|obj+0| using 256 processes -- indicates that \ddo's organizatorial overhead is much smaller than that of DistArray.
Supporting evidence for this is that the initialization of an empty DistArray (\verb|copy_empty|) becomes disproportionately costly when increasing the number of processes used.

\subsection{Strong Scaling: Real-World Application Speedup -- the Wiener filter}\label{sec:wiener}
\ddo\ was initially developed for \nifty\cite{nifty},
    a library for building signal inference algorithms in the framework of \emph{information field theory} (IFT)\cite{Ensslin:2008iu}.
Within \nifty~v2 all of the parallelization is done via \ddo;
    the code of \nifty\ itself is almost completely agnostic of the parallelization and completely agnostic of MPI.\\
A standard computational operation in IFT-based signal reconstruction is the \emph{Wiener filter}\cite{wiener1949}.
For this performance test, we use a Wiener filter implemented in \nifty\ to reconstruct the realization of a Gaussian random field -- called the \emph{signal}.
Assume we performed a hypothetical measurement which produced some \emph{data}.
The data model is
\begin{equation}
\mathrm{data} = R(\mathrm{signal}) + \mathrm{noise}
\end{equation}
where $R$ is a smoothing operator and \emph{noise} is additive Gaussian noise.
In figure \ref{fig_wiener_filter} one sees the three steps:
\begin{itemize}
    \item the true signal we try to reconstruct,
    \item the data one gets from the hypothetical measurement, and
    \item the reconstructed signal field that according to the Wiener filter has most likely produced the $data$.
\end{itemize}
Table \ref{tab_wiener_filter} shows the scaling behavior of the reconstruction code,
    run with a resolution of $8192 \times 8192$.
Here, $n$ is the number of used processes and $t_n$ the respective execution time.
The relative speedup $s_n = 2t_2/t_n$\footnote{Since the combination of NIFTy and pyfftw exhibits an unexpected speed malus for one process, we chose the two-process timing as the benchmark's baseline.}
    is the ratio of execution times: parallel versus serial.
In the case of exactly linear scaling $s_n$ is equal to $n$.
Furthermore we define the scaling quality $q = 1/(1+\log(n/s_n))$,
    which compares $s_n$ with linear scaling in terms of orders of magnitude.
A value $q=1$ represents linear scaling and $q \ge 1$ super-linear scaling.
\begin{table}[h!]
\caption{Execution time scaling of a Wiener filter reconstruction on a grid of size $8192 \times 8192$.}\label{tab_wiener_filter}
      \begin{tabular}{c|ccccccccc}
            $\#\mathrm{nodes}$ & 1  & 1  & 1 & 1 & 2 & 4 & 8 & 16 & 32 \\
            $\#\mathrm{processes:~} n$ & 1  & 2  & 3 & 4 & 8 & 16 & 32 & 64 & 128 \\ \hline
            $t [\SI{}{\second}]$ & 1618 & 622.0 & 404.2 & 364.2 & 181.7 & 94.50 & 46.79 & 18.74 & 8.56 \\
            $s_n = 2t_2/t_n$ & 0.769 &  2.00 &  3.08 &  3.42 &  6.85 &  13.2 &  26.6 &  66.4 &  145\\
            $q_n = \scriptstyle{1/(1+\log(\frac{n}{s_n}))}$ & 0.900 &  1.00 &  1.01 &  0.94 &  0.94 &  0.92 &  0.93 &  1.02 &  1.06
      \end{tabular}
\end{table}

This benchmark illustrates that even in a real-life application super-linear scaling is possible to achieve for a sufficiently large number of processes.
This is due to the operations that are needed in order to perform the Wiener filtering:
    basic point-wise arithmetics that do not involve any inter-process communication and
    Fourier transformations that are handled by the high-performance library \emph{FFTW}\cite{FFTW05}.
While the problem size remains constant, the amount of available CPU cache increases with the number of processes,
    which explains the super-linear scaling,
    cf.\ section \ref{sec:strong_scaling}.

\section{Summary \& Outlook}\label{sec:summary}
We introduced \ddo, a Python module for cluster-distributed multi-dimensional numerical arrays.
It can be understood as a layer of abstraction between abstract algorithm code and actual data-distribution logic.
We argued why we developed \ddo\ as a package following a low-level parallelization ansatz
    and why we built it on MPI.
Compared to other packages available for data parallelization, \ddo\ has the advantage
    of being ready for action on one as well as several hundreds of CPUs,
    of being highly portable and customizable as it is built with Python,
    that it is faster in many circumstances,
    and that it is able to treat arrays of arbitrary dimension.

For the future, we plan to cover more of \emph{numpy}'s interface such that working with \ddo\ becomes even more convenient.
Furthermore we evaluate the option to build a \ddo\ distributor in order to support \emph{scalapy}'s block-cyclic distribution strategy directly.
This will open up a whole new class of applications \ddo\ then can be used for.

\ddo\ is open source software licensed under the GNU General Public License v3 (GPL-3) and is available by \url{https://gitlab.mpcdf.mpg.de/ift/D2O}.


\begin{backmatter}

\section*{Competing interests}
  The authors declare that they have no competing interests.

\section*{Author's contributions}
TS is the principal researcher for the work proposed in this article.
His contributions include the primal idea, the implementation of presented software package, the conduction of the performance tests and the writing of the article.
MG, FB and TE helped working out the conceptual structure of the software package and drafting the manuscript.
FB also played a pivotal role for executing the performance tests.
Furthermore, TE also fulfilled the role of the principal investigator.
All authors read and approved the final manuscript.

\section*{Acknowledgments}
We want to thank Jait Dixit, Philipp Franck, Reimar Leike, Fotis Megas, Martin Reinecke and Csongor Varady for useful discussions and support.
We acknowledge the support by the DFG Cluster of Excellence "Origin and Structure of the Universe" and the Studienstiftung des deutschen Volkes.
The performance tests have been carried out on the computing facilities of the Computational Center for Particle and Astrophysics (C2PAP).
We are grateful for the support by Dr. Alexey Krukau through the Computational Center for Particle and Astrophysics (C2PAP).


\bibliographystyle{bmc-mathphys} 
\bibliography{bib}


\begin{thebibliography}{27}
\ifx \bisbn   \undefined \def \bisbn  #1{ISBN #1}\fi
\ifx \binits  \undefined \def \binits#1{#1}\fi
\ifx \bauthor  \undefined \def \bauthor#1{#1}\fi
\ifx \batitle  \undefined \def \batitle#1{#1}\fi
\ifx \bjtitle  \undefined \def \bjtitle#1{#1}\fi
\ifx \bvolume  \undefined \def \bvolume#1{\textbf{#1}}\fi
\ifx \byear  \undefined \def \byear#1{#1}\fi
\ifx \bissue  \undefined \def \bissue#1{#1}\fi
\ifx \bfpage  \undefined \def \bfpage#1{#1}\fi
\ifx \blpage  \undefined \def \blpage #1{#1}\fi
\ifx \burl  \undefined \def \burl#1{\textsf{#1}}\fi
\ifx \doiurl  \undefined \def \doiurl#1{\textsf{#1}}\fi
\ifx \betal  \undefined \def \betal{\textit{et al.}}\fi
\ifx \binstitute  \undefined \def \binstitute#1{#1}\fi
\ifx \binstitutionaled  \undefined \def \binstitutionaled#1{#1}\fi
\ifx \bctitle  \undefined \def \bctitle#1{#1}\fi
\ifx \beditor  \undefined \def \beditor#1{#1}\fi
\ifx \bpublisher  \undefined \def \bpublisher#1{#1}\fi
\ifx \bbtitle  \undefined \def \bbtitle#1{#1}\fi
\ifx \bedition  \undefined \def \bedition#1{#1}\fi
\ifx \bseriesno  \undefined \def \bseriesno#1{#1}\fi
\ifx \blocation  \undefined \def \blocation#1{#1}\fi
\ifx \bsertitle  \undefined \def \bsertitle#1{#1}\fi
\ifx \bsnm \undefined \def \bsnm#1{#1}\fi
\ifx \bsuffix \undefined \def \bsuffix#1{#1}\fi
\ifx \bparticle \undefined \def \bparticle#1{#1}\fi
\ifx \barticle \undefined \def \barticle#1{#1}\fi
\ifx \bconfdate \undefined \def \bconfdate #1{#1}\fi
\ifx \botherref \undefined \def \botherref #1{#1}\fi
\ifx \url \undefined \def \url#1{\textsf{#1}}\fi
\ifx \bchapter \undefined \def \bchapter#1{#1}\fi
\ifx \bbook \undefined \def \bbook#1{#1}\fi
\ifx \bcomment \undefined \def \bcomment#1{#1}\fi
\ifx \oauthor \undefined \def \oauthor#1{#1}\fi
\ifx \citeauthoryear \undefined \def \citeauthoryear#1{#1}\fi
\ifx \endbibitem  \undefined \def \endbibitem {}\fi
\ifx \bconflocation  \undefined \def \bconflocation#1{#1}\fi
\ifx \arxivurl  \undefined \def \arxivurl#1{\textsf{#1}}\fi
\csname PreBibitemsHook\endcsname

\bibitem{greiner.electron_density}
\begin{botherref}
\oauthor{\bsnm{{Greiner}}, \binits{M.}},
\oauthor{\bsnm{{Schnitzeler}}, \binits{D.H.F.M.}},
\oauthor{\bsnm{{Ensslin}}, \binits{T.A.}}:
{Tomography of the Galactic free electron density with the Square Kilometer
  Array}.
ArXiv e-prints
(2015).
\arxivurl{1512.03480}
\end{botherref}
\endbibitem

\bibitem{junklewitz.resolve}
\begin{barticle}
\bauthor{\bsnm{{Junklewitz}}, \binits{H.}},
\bauthor{\bsnm{{Bell}}, \binits{M.R.}},
\bauthor{\bsnm{{Selig}}, \binits{M.}},
\bauthor{\bsnm{{En{\ss}lin}}, \binits{T.A.}}:
\batitle{{RESOLVE: A new algorithm for aperture synthesis imaging of extended
  emission in radio astronomy}}.
\bjtitle{\aap}
\bvolume{586},
\bfpage{76}
(\byear{2016}).
doi:\doiurl{10.1051/0004-6361/201323094}.
\arxivurl{1311.5282}
\end{barticle}
\endbibitem

\bibitem{nifty}
\begin{barticle}
\bauthor{\bsnm{{Selig}}, \binits{M.}},
\bauthor{\bsnm{{Bell}}, \binits{M.R.}},
\bauthor{\bsnm{{Junklewitz}}, \binits{H.}},
\bauthor{\bsnm{{Oppermann}}, \binits{N.}},
\bauthor{\bsnm{{Reinecke}}, \binits{M.}},
\bauthor{\bsnm{{Greiner}}, \binits{M.}},
\bauthor{\bsnm{{Pachajoa}}, \binits{C.}},
\bauthor{\bsnm{{En{\ss}lin}}, \binits{T.A.}}:
\batitle{{NIFTY - Numerical Information Field Theory. A versatile PYTHON
  library for signal inference}}.
\bjtitle{\aap}
\bvolume{554},
\bfpage{26}
(\byear{2013}).
doi:\doiurl{10.1051/0004-6361/201321236}.
\arxivurl{1301.4499}
\end{barticle}
\endbibitem

\bibitem{walt.numpy}
\begin{barticle}
\bauthor{\bparticle{van~der} \bsnm{Walt}, \binits{S.}},
\bauthor{\bsnm{Colbert}, \binits{S.C.}},
\bauthor{\bsnm{Varoquaux}, \binits{G.}}:
\batitle{The numpy array: A structure for efficient numerical computation}.
\bjtitle{Computing in Science and Engineering}
\bvolume{13}(\bissue{2}),
\bfpage{22}--\blpage{30}
(\byear{2011}).
doi:\doiurl{10.1109/MCSE.2011.37}
\end{barticle}
\endbibitem

\bibitem{mpi-1-standard}
\begin{barticle}
\bauthor{\bsnm{Forum}, \binits{M.P.I.}}:
\batitle{{MPI}: A message passing interface standard}.
\bjtitle{International Journal of Supercomputer Applications}
\bvolume{8}(\bissue{3/4}),
\bfpage{159}--\blpage{416}
(\byear{1994})
\end{barticle}
\endbibitem

\bibitem{mpi-2-standard}
\begin{barticle}
\bauthor{\bsnm{{Message Passing Interface Forum}}}:
\batitle{{MPI2}: A message passing interface standard}.
\bjtitle{High Performance Computing Applications}
\bvolume{12}(\bissue{1--2}),
\bfpage{1}--\blpage{299}
(\byear{1998})
\end{barticle}
\endbibitem

\bibitem{distarray}
\begin{botherref}
\oauthor{\bsnm{Enthought}, \binits{I.}}:
DistArray: Think globally, act locally
(2016).
\url{http://docs.enthought.com/distarray/}
Accessed 2016-03-24
\end{botherref}
\endbibitem

\bibitem{Frigo:1999:FFT:301618.301661}
\begin{bchapter}
\bauthor{\bsnm{Frigo}, \binits{M.}}:
\bctitle{A fast fourier transform compiler}.
In: \bbtitle{Proceedings of the ACM SIGPLAN 1999 Conference on Programming
  Language Design and Implementation}.
\bsertitle{PLDI '99},
pp. \bfpage{169}--\blpage{180}.
\bpublisher{ACM},
\blocation{New York, NY, USA}
(\byear{1999}).
doi:\doiurl{10.1145/301618.301661}.
\burl{http://doi.acm.org/10.1145/301618.301661}
\end{bchapter}
\endbibitem

\bibitem{slug}
\begin{bbook}
\bauthor{\bsnm{Blackford}, \binits{L.S.}},
\bauthor{\bsnm{Choi}, \binits{J.}},
\bauthor{\bsnm{Cleary}, \binits{A.}},
\bauthor{\bsnm{D'Azevedo}, \binits{E.}},
\bauthor{\bsnm{Demmel}, \binits{J.}},
\bauthor{\bsnm{Dhillon}, \binits{I.}},
\bauthor{\bsnm{Dongarra}, \binits{J.}},
\bauthor{\bsnm{Hammarling}, \binits{S.}},
\bauthor{\bsnm{Henry}, \binits{G.}},
\bauthor{\bsnm{Petitet}, \binits{A.}},
\bauthor{\bsnm{Stanley}, \binits{K.}},
\bauthor{\bsnm{Walker}, \binits{D.}},
\bauthor{\bsnm{Whaley}, \binits{R.C.}}:
\bbtitle{{ScaLAPACK} Users' Guide}.
\bpublisher{Society for Industrial and Applied Mathematics},
\blocation{Philadelphia, PA}
(\byear{1997})
\end{bbook}
\endbibitem

\bibitem{DadoneGhostCellGrid}
\begin{barticle}
\bauthor{\bsnm{{Dadone}}, \binits{A.}},
\bauthor{\bsnm{{Grossman}}, \binits{B.}}:
\batitle{{Ghost-Cell Method for Inviscid Two-Dimensional Flows on Cartesian
  Grids}}.
\bjtitle{AIAA Journal}
\bvolume{42},
\bfpage{2499}--\blpage{2507}
(\byear{2004}).
doi:\doiurl{10.2514/1.697}
\end{barticle}
\endbibitem

\bibitem{PER-GRA:2007.ipython}
\begin{barticle}
\bauthor{\bsnm{P\'erez}, \binits{F.}},
\bauthor{\bsnm{Granger}, \binits{B.E.}}:
\batitle{{IP}ython: a system for interactive scientific computing}.
\bjtitle{Computing in Science and Engineering}
\bvolume{9}(\bissue{3}),
\bfpage{21}--\blpage{29}
(\byear{2007}).
doi:\doiurl{10.1109/MCSE.2007.53}
\end{barticle}
\endbibitem

\bibitem{scalapack.online}
\begin{botherref}
\oauthor{\bsnm{Team}, \binits{S.}}:
ScaLAPACK Web Page
(2016).
\url{www.netlib.org/scalapack/}
Accessed 2016-03-23
\end{botherref}
\endbibitem

\bibitem{petsc-web-page}
\begin{botherref}
\oauthor{\bsnm{Balay}, \binits{S.}},
\oauthor{\bsnm{Abhyankar}, \binits{S.}},
\oauthor{\bsnm{Adams}, \binits{M.F.}},
\oauthor{\bsnm{Brown}, \binits{J.}},
\oauthor{\bsnm{Brune}, \binits{P.}},
\oauthor{\bsnm{Buschelman}, \binits{K.}},
\oauthor{\bsnm{Dalcin}, \binits{L.}},
\oauthor{\bsnm{Eijkhout}, \binits{V.}},
\oauthor{\bsnm{Gropp}, \binits{W.D.}},
\oauthor{\bsnm{Kaushik}, \binits{D.}},
\oauthor{\bsnm{Knepley}, \binits{M.G.}},
\oauthor{\bsnm{McInnes}, \binits{L.C.}},
\oauthor{\bsnm{Rupp}, \binits{K.}},
\oauthor{\bsnm{Smith}, \binits{B.F.}},
\oauthor{\bsnm{Zampini}, \binits{S.}},
\oauthor{\bsnm{Zhang}, \binits{H.}}:
{PETS}c {W}eb page.
\url{http://www.mcs.anl.gov/petsc}
(2015).
\url{http://www.mcs.anl.gov/petsc}
\end{botherref}
\endbibitem

\bibitem{mckerns.pathos}
\begin{botherref}
\oauthor{\bsnm{McKerns}, \binits{M.M.}},
\oauthor{\bsnm{Strand}, \binits{L.}},
\oauthor{\bsnm{Sullivan}, \binits{T.}},
\oauthor{\bsnm{Fang}, \binits{A.}},
\oauthor{\bsnm{Aivazis}, \binits{M.A.G.}}:
Building a framework for predictive science.
CoRR
\textbf{abs/1202.1056}
(2012)
\end{botherref}
\endbibitem

\bibitem{TOP500}
\begin{botherref}
\oauthor{\bsnm{Strohmaier}, \binits{E.}},
\oauthor{\bsnm{Dongarra}, \binits{J.}},
\oauthor{\bsnm{Simon}, \binits{H.}},
\oauthor{\bsnm{Meuer}, \binits{M.}}:
The TOP500 project
(2015).
\url{http://www.top500.org/lists/2015/11/}
Accessed 2016-03-24
\end{botherref}
\endbibitem

\bibitem{Zaharia:2010:SCC:1863103.1863113}
\begin{bchapter}
\bauthor{\bsnm{Zaharia}, \binits{M.}},
\bauthor{\bsnm{Chowdhury}, \binits{M.}},
\bauthor{\bsnm{Franklin}, \binits{M.J.}},
\bauthor{\bsnm{Shenker}, \binits{S.}},
\bauthor{\bsnm{Stoica}, \binits{I.}}:
\bctitle{Spark: Cluster computing with working sets}.
In: \bbtitle{Proceedings of the 2Nd USENIX Conference on Hot Topics in Cloud
  Computing}.
\bsertitle{HotCloud'10},
pp. \bfpage{10}--\blpage{10}.
\bpublisher{USENIX Association},
\blocation{Berkeley, CA, USA}
(\byear{2010}).
\burl{http://dl.acm.org/citation.cfm?id=1863103.1863113}
\end{bchapter}
\endbibitem

\bibitem{hadoop}
\begin{botherref}
\oauthor{\bsnm{{Apache Software Foundation}}}:
Hadoop
(2016).
\url{https://hadoop.apache.org}
Accessed 2016-03-23
\end{botherref}
\endbibitem

\bibitem{gabriel04:_open_mpi}
\begin{bchapter}
\bauthor{\bsnm{Gabriel}, \binits{E.}},
\bauthor{\bsnm{Fagg}, \binits{G.E.}},
\bauthor{\bsnm{Bosilca}, \binits{G.}},
\bauthor{\bsnm{Angskun}, \binits{T.}},
\bauthor{\bsnm{Dongarra}, \binits{J.J.}},
\bauthor{\bsnm{Squyres}, \binits{J.M.}},
\bauthor{\bsnm{Sahay}, \binits{V.}},
\bauthor{\bsnm{Kambadur}, \binits{P.}},
\bauthor{\bsnm{Barrett}, \binits{B.}},
\bauthor{\bsnm{Lumsdaine}, \binits{A.}},
\bauthor{\bsnm{Castain}, \binits{R.H.}},
\bauthor{\bsnm{Daniel}, \binits{D.J.}},
\bauthor{\bsnm{Graham}, \binits{R.L.}},
\bauthor{\bsnm{Woodall}, \binits{T.S.}}:
\bctitle{Open {MPI}: Goals, concept, and design of a next generation {MPI}
  implementation}.
In: \bbtitle{Proceedings, 11th European PVM/MPI Users' Group Meeting},
\bconflocation{Budapest, Hungary},
pp. \bfpage{97}--\blpage{104}
(\byear{2004})
\end{bchapter}
\endbibitem

\bibitem{mpich2}
\begin{botherref}
\oauthor{\bsnm{Team}, \binits{M.}}:
MPICH2: High-Performance Portable MPI
(2016).
\url{www.mcs.anl.gov/mpich2}
Accessed 2016-03-24
\end{botherref}
\endbibitem

\bibitem{intelmpi}
\begin{botherref}
\oauthor{\bsnm{Corporation}, \binits{I.}}:
Intel MPI Library
(2016).
\url{https://software.intel.com/en-us/intel-mpi-library}
Accessed 2016-06-06
\end{botherref}
\endbibitem

\bibitem{Dalcin20051108.mpi4py}
\begin{barticle}
\bauthor{\bsnm{Dalcín}, \binits{L.}},
\bauthor{\bsnm{Paz}, \binits{R.}},
\bauthor{\bsnm{Storti}, \binits{M.}}:
\batitle{\{MPI\} for python}.
\bjtitle{Journal of Parallel and Distributed Computing}
\bvolume{65}(\bissue{9}),
\bfpage{1108}--\blpage{1115}
(\byear{2005}).
doi:\doiurl{10.1016/j.jpdc.2005.03.010}
\end{barticle}
\endbibitem

\bibitem{pyfftw}
\begin{botherref}
\oauthor{\bsnm{Gomersall}, \binits{H.}}:
pyFFTW: a pythonic wrapper around FFTW.
We use the mpi branch available at \url{https://github.com/fredRos/pyFFTW}
(2016).
\url{https://hgomersall.github.io/pyFFTW}
Accessed 2016-03-23
\end{botherref}
\endbibitem

\bibitem{c2pap}
\begin{botherref}
\oauthor{\bsnm{Universe}, \binits{E.C.}}:
Excellence Cluster Universe
(2016).
\url{http://www.universe-cluster.de/c2pap}
Accessed 2016-04-06
\end{botherref}
\endbibitem

\bibitem{Ensslin:2008iu}
\begin{barticle}
\bauthor{\bsnm{En{\ss}lin}, \binits{T.A.}},
\bauthor{\bsnm{Frommert}, \binits{M.}},
\bauthor{\bsnm{Kitaura}, \binits{F.S.}}:
\batitle{{Information field theory for cosmological perturbation reconstruction
  and non-linear signal analysis}}.
\bjtitle{Phys. Rev.}
\bvolume{D80},
\bfpage{105005}
(\byear{2009}).
doi:\doiurl{10.1103/PhysRevD.80.105005}.
\arxivurl{0806.3474}
\end{barticle}
\endbibitem

\bibitem{wiener1949}
\begin{bbook}
\bauthor{\bsnm{Wiener}, \binits{N.}}:
\bbtitle{Extrapolation, Interpolation and Smoothing of Stationary Time Series,
  with Engineering Applications}.
\bpublisher{Technology Press and Wiley},
\blocation{New York}
(\byear{1949}).
\bcomment{note: Originally issued in Feb 1942 as a classified Nat. Defense Res.
  Council Rep.}
\end{bbook}
\endbibitem

\bibitem{FFTW05}
\begin{barticle}
\bauthor{\bsnm{Frigo}, \binits{M.}},
\bauthor{\bsnm{Johnson}, \binits{S.G.}}:
\batitle{The design and implementation of {FFTW3}}.
\bjtitle{Proceedings of the IEEE}
\bvolume{93}(\bissue{2}),
\bfpage{216}--\blpage{231}
(\byear{2005}).
\bcomment{Special issue on ``Program Generation, Optimization, and Platform
  Adaptation''}
\end{barticle}
\endbibitem

\bibitem{python_iterators}
\begin{botherref}
\oauthor{\bsnm{Ka-Ping~Yee}, \binits{G.v.R.}}:
PEP 234 -- Iterators
(2016).
\url{https://www.python.org/dev/peps/pep-0234/}
Accessed 2016-04-12
\end{botherref}
\endbibitem

\end{thebibliography}

\newcommand{\BMCxmlcomment}[1]{}

\BMCxmlcomment{

<refgrp>

<bibl id="B1">
  <title><p>{Tomography of the Galactic free electron density with the Square
  Kilometer Array}</p></title>
  <aug>
    <au><snm>{Greiner}</snm><fnm>M.</fnm></au>
    <au><snm>{Schnitzeler}</snm><fnm>D. H. F. M.</fnm></au>
    <au><snm>{Ensslin}</snm><fnm>T. A.</fnm></au>
  </aug>
  <source>ArXiv e-prints</source>
  <pubdate>2015</pubdate>
</bibl>

<bibl id="B2">
  <title><p>{RESOLVE: A new algorithm for aperture synthesis imaging of
  extended emission in radio astronomy}</p></title>
  <aug>
    <au><snm>{Junklewitz}</snm><fnm>H.</fnm></au>
    <au><snm>{Bell}</snm><fnm>M. R.</fnm></au>
    <au><snm>{Selig}</snm><fnm>M.</fnm></au>
    <au><snm>{En{\ss}lin}</snm><fnm>T. A.</fnm></au>
  </aug>
  <source>\aap</source>
  <pubdate>2016</pubdate>
  <volume>586</volume>
  <fpage>A76</fpage>
</bibl>

<bibl id="B3">
  <title><p>{NIFTY - Numerical Information Field Theory. A versatile PYTHON
  library for signal inference}</p></title>
  <aug>
    <au><snm>{Selig}</snm><fnm>M.</fnm></au>
    <au><snm>{Bell}</snm><fnm>M. R.</fnm></au>
    <au><snm>{Junklewitz}</snm><fnm>H.</fnm></au>
    <au><snm>{Oppermann}</snm><fnm>N.</fnm></au>
    <au><snm>{Reinecke}</snm><fnm>M.</fnm></au>
    <au><snm>{Greiner}</snm><fnm>M.</fnm></au>
    <au><snm>{Pachajoa}</snm><fnm>C.</fnm></au>
    <au><snm>{En{\ss}lin}</snm><fnm>T. A.</fnm></au>
  </aug>
  <source>\aap</source>
  <pubdate>2013</pubdate>
  <volume>554</volume>
  <fpage>A26</fpage>
</bibl>

<bibl id="B4">
  <title><p>The NumPy Array: A Structure for Efficient Numerical
  Computation</p></title>
  <aug>
    <au><snm>Walt</snm><fnm>S</fnm></au>
    <au><snm>Colbert</snm><fnm>SC</fnm></au>
    <au><snm>Varoquaux</snm><fnm>G</fnm></au>
  </aug>
  <source>Computing in Science and Engineering</source>
  <publisher>Los Alamitos, CA, USA: IEEE Computer Society</publisher>
  <pubdate>2011</pubdate>
  <volume>13</volume>
  <issue>2</issue>
  <fpage>22</fpage>
  <lpage>30</lpage>
</bibl>

<bibl id="B5">
  <title><p>{MPI}: A Message Passing Interface Standard</p></title>
  <aug>
    <au><snm>Forum</snm><fnm>MPI</fnm></au>
  </aug>
  <source>International Journal of Supercomputer Applications</source>
  <pubdate>1994</pubdate>
  <volume>8</volume>
  <issue>3/4</issue>
  <fpage>159</fpage>
  <lpage>-416</lpage>
</bibl>

<bibl id="B6">
  <title><p>{MPI2}: A Message Passing Interface Standard</p></title>
  <aug>
    <au><cnm>{Message Passing Interface Forum}</cnm></au>
  </aug>
  <source>High Performance Computing Applications</source>
  <pubdate>1998</pubdate>
  <volume>12</volume>
  <issue>1--2</issue>
  <fpage>1</fpage>
  <lpage>-299</lpage>
</bibl>

<bibl id="B7">
  <title><p>DistArray: Think globally, act locally</p></title>
  <aug>
    <au><snm>Enthought</snm><fnm>I</fnm></au>
  </aug>
  <pubdate>2016</pubdate>
  <url>http://docs.enthought.com/distarray/</url>
</bibl>

<bibl id="B8">
  <title><p>A Fast Fourier Transform Compiler</p></title>
  <aug>
    <au><snm>Frigo</snm><fnm>M</fnm></au>
  </aug>
  <source>Proceedings of the ACM SIGPLAN 1999 Conference on Programming
  Language Design and Implementation</source>
  <publisher>New York, NY, USA: ACM</publisher>
  <series><title><p>PLDI '99</p></title></series>
  <pubdate>1999</pubdate>
  <fpage>169</fpage>
  <lpage>-180</lpage>
  <url>http://doi.acm.org/10.1145/301618.301661</url>
</bibl>

<bibl id="B9">
  <title><p>{ScaLAPACK} Users' Guide</p></title>
  <aug>
    <au><snm>Blackford</snm><fnm>L. S.</fnm></au>
    <au><snm>Choi</snm><fnm>J.</fnm></au>
    <au><snm>Cleary</snm><fnm>A.</fnm></au>
    <au><snm>D'Azevedo</snm><fnm>E.</fnm></au>
    <au><snm>Demmel</snm><fnm>J.</fnm></au>
    <au><snm>Dhillon</snm><fnm>I.</fnm></au>
    <au><snm>Dongarra</snm><fnm>J.</fnm></au>
    <au><snm>Hammarling</snm><fnm>S.</fnm></au>
    <au><snm>Henry</snm><fnm>G.</fnm></au>
    <au><snm>Petitet</snm><fnm>A.</fnm></au>
    <au><snm>Stanley</snm><fnm>K.</fnm></au>
    <au><snm>Walker</snm><fnm>D.</fnm></au>
    <au><snm>Whaley</snm><fnm>R. C.</fnm></au>
  </aug>
  <publisher>Philadelphia, PA: Society for Industrial and Applied
  Mathematics</publisher>
  <pubdate>1997</pubdate>
</bibl>

<bibl id="B10">
  <title><p>{Ghost-Cell Method for Inviscid Two-Dimensional Flows on Cartesian
  Grids}</p></title>
  <aug>
    <au><snm>{Dadone}</snm><fnm>A.</fnm></au>
    <au><snm>{Grossman}</snm><fnm>B.</fnm></au>
  </aug>
  <source>AIAA Journal</source>
  <pubdate>2004</pubdate>
  <volume>42</volume>
  <fpage>2499</fpage>
  <lpage>2507</lpage>
</bibl>

<bibl id="B11">
  <title><p>{IP}ython: a System for Interactive Scientific
  Computing</p></title>
  <aug>
    <au><snm>P\'erez</snm><fnm>F</fnm></au>
    <au><snm>Granger</snm><fnm>BE</fnm></au>
  </aug>
  <source>Computing in Science and Engineering</source>
  <publisher>IEEE Computer Society</publisher>
  <pubdate>2007</pubdate>
  <volume>9</volume>
  <issue>3</issue>
  <fpage>21</fpage>
  <lpage>-29</lpage>
  <url>http://ipython.org</url>
</bibl>

<bibl id="B12">
  <title><p>ScaLAPACK Web Page</p></title>
  <aug>
    <au><snm>Team</snm><fnm>S</fnm></au>
  </aug>
  <pubdate>2016</pubdate>
  <url>www.netlib.org/scalapack/</url>
</bibl>

<bibl id="B13">
  <title><p>{PETS}c {W}eb page</p></title>
  <aug>
    <au><snm>Balay</snm><fnm>S</fnm></au>
    <au><snm>Abhyankar</snm><fnm>S</fnm></au>
    <au><snm>Adams</snm><fnm>MF</fnm></au>
    <au><snm>Brown</snm><fnm>J</fnm></au>
    <au><snm>Brune</snm><fnm>P</fnm></au>
    <au><snm>Buschelman</snm><fnm>K</fnm></au>
    <au><snm>Dalcin</snm><fnm>L</fnm></au>
    <au><snm>Eijkhout</snm><fnm>V</fnm></au>
    <au><snm>Gropp</snm><fnm>WD</fnm></au>
    <au><snm>Kaushik</snm><fnm>D</fnm></au>
    <au><snm>Knepley</snm><fnm>MG</fnm></au>
    <au><snm>McInnes</snm><fnm>LC</fnm></au>
    <au><snm>Rupp</snm><fnm>K</fnm></au>
    <au><snm>Smith</snm><fnm>BF</fnm></au>
    <au><snm>Zampini</snm><fnm>S</fnm></au>
    <au><snm>Zhang</snm><fnm>H</fnm></au>
  </aug>
  <source>\url{http://www.mcs.anl.gov/petsc}</source>
  <pubdate>2015</pubdate>
  <url>http://www.mcs.anl.gov/petsc</url>
</bibl>

<bibl id="B14">
  <title><p>Building a Framework for Predictive Science</p></title>
  <aug>
    <au><snm>McKerns</snm><fnm>MM</fnm></au>
    <au><snm>Strand</snm><fnm>L</fnm></au>
    <au><snm>Sullivan</snm><fnm>T</fnm></au>
    <au><snm>Fang</snm><fnm>A</fnm></au>
    <au><snm>Aivazis</snm><fnm>MAG</fnm></au>
  </aug>
  <source>CoRR</source>
  <pubdate>2012</pubdate>
  <volume>abs/1202.1056</volume>
  <url>http://arxiv.org/abs/1202.1056</url>
</bibl>

<bibl id="B15">
  <title><p>The TOP500 project</p></title>
  <aug>
    <au><snm>Strohmaier</snm><fnm>E</fnm></au>
    <au><snm>Dongarra</snm><fnm>J</fnm></au>
    <au><snm>Simon</snm><fnm>H</fnm></au>
    <au><snm>Meuer</snm><fnm>M</fnm></au>
  </aug>
  <pubdate>2015</pubdate>
  <url>http://www.top500.org/lists/2015/11/</url>
</bibl>

<bibl id="B16">
  <title><p>Spark: Cluster Computing with Working Sets</p></title>
  <aug>
    <au><snm>Zaharia</snm><fnm>M</fnm></au>
    <au><snm>Chowdhury</snm><fnm>M</fnm></au>
    <au><snm>Franklin</snm><fnm>MJ</fnm></au>
    <au><snm>Shenker</snm><fnm>S</fnm></au>
    <au><snm>Stoica</snm><fnm>I</fnm></au>
  </aug>
  <source>Proceedings of the 2Nd USENIX Conference on Hot Topics in Cloud
  Computing</source>
  <publisher>Berkeley, CA, USA: USENIX Association</publisher>
  <series><title><p>HotCloud'10</p></title></series>
  <pubdate>2010</pubdate>
  <fpage>10</fpage>
  <lpage>-10</lpage>
  <url>http://dl.acm.org/citation.cfm?id=1863103.1863113</url>
</bibl>

<bibl id="B17">
  <title><p>Hadoop</p></title>
  <aug>
    <au><cnm>{Apache Software Foundation}</cnm></au>
  </aug>
  <pubdate>2016</pubdate>
  <url>https://hadoop.apache.org</url>
</bibl>

<bibl id="B18">
  <title><p>Open {MPI}: Goals, Concept, and Design of a Next Generation {MPI}
  Implementation</p></title>
  <aug>
    <au><snm>Gabriel</snm><fnm>E</fnm></au>
    <au><snm>Fagg</snm><fnm>GE</fnm></au>
    <au><snm>Bosilca</snm><fnm>G</fnm></au>
    <au><snm>Angskun</snm><fnm>T</fnm></au>
    <au><snm>Dongarra</snm><fnm>JJ</fnm></au>
    <au><snm>Squyres</snm><fnm>JM</fnm></au>
    <au><snm>Sahay</snm><fnm>V</fnm></au>
    <au><snm>Kambadur</snm><fnm>P</fnm></au>
    <au><snm>Barrett</snm><fnm>B</fnm></au>
    <au><snm>Lumsdaine</snm><fnm>A</fnm></au>
    <au><snm>Castain</snm><fnm>RH</fnm></au>
    <au><snm>Daniel</snm><fnm>DJ</fnm></au>
    <au><snm>Graham</snm><fnm>RL</fnm></au>
    <au><snm>Woodall</snm><fnm>TS</fnm></au>
  </aug>
  <source>Proceedings, 11th European PVM/MPI Users' Group Meeting</source>
  <publisher>Budapest, Hungary</publisher>
  <pubdate>2004</pubdate>
  <fpage>97</fpage>
  <lpage>-104</lpage>
</bibl>

<bibl id="B19">
  <title><p>MPICH2: High-Performance Portable MPI</p></title>
  <aug>
    <au><snm>Team</snm><fnm>MPICH</fnm></au>
  </aug>
  <pubdate>2016</pubdate>
  <url>www.mcs.anl.gov/mpich2</url>
</bibl>

<bibl id="B20">
  <title><p>Intel MPI Library</p></title>
  <aug>
    <au><snm>Corporation</snm><fnm>I</fnm></au>
  </aug>
  <pubdate>2016</pubdate>
  <url>https://software.intel.com/en-us/intel-mpi-library</url>
</bibl>

<bibl id="B21">
  <title><p>\{MPI\} for Python</p></title>
  <aug>
    <au><snm>Dalcín</snm><fnm>L</fnm></au>
    <au><snm>Paz</snm><fnm>R</fnm></au>
    <au><snm>Storti</snm><fnm>M</fnm></au>
  </aug>
  <source>Journal of Parallel and Distributed Computing</source>
  <pubdate>2005</pubdate>
  <volume>65</volume>
  <issue>9</issue>
  <fpage>1108</fpage>
  <lpage>1115</lpage>
  <url>http://www.sciencedirect.com/science/article/pii/S0743731505000560</url>
</bibl>

<bibl id="B22">
  <title><p>pyFFTW: a pythonic wrapper around FFTW</p></title>
  <aug>
    <au><snm>Gomersall</snm><fnm>H</fnm></au>
  </aug>
  <pubdate>2016</pubdate>
  <url>https://hgomersall.github.io/pyFFTW</url>
  <note>We use the mpi branch available at
  \url{https://github.com/fredRos/pyFFTW}</note>
</bibl>

<bibl id="B23">
  <title><p>Excellence Cluster Universe</p></title>
  <aug>
    <au><snm>Universe</snm><fnm>EC</fnm></au>
  </aug>
  <pubdate>2016</pubdate>
  <url>http://www.universe-cluster.de/c2pap</url>
</bibl>

<bibl id="B24">
  <title><p>{Information field theory for cosmological perturbation
  reconstruction and non-linear signal analysis}</p></title>
  <aug>
    <au><snm>En{\ss}lin</snm><fnm>TA</fnm></au>
    <au><snm>Frommert</snm><fnm>M</fnm></au>
    <au><snm>Kitaura</snm><fnm>FS</fnm></au>
  </aug>
  <source>Phys. Rev.</source>
  <pubdate>2009</pubdate>
  <volume>D80</volume>
  <fpage>105005</fpage>
</bibl>

<bibl id="B25">
  <title><p>Extrapolation, Interpolation and Smoothing of Stationary Time
  Series, with Engineering Applications</p></title>
  <aug>
    <au><snm>Wiener</snm><fnm>N.</fnm></au>
  </aug>
  <publisher>New York: Technology Press and Wiley</publisher>
  <pubdate>1949</pubdate>
  <note>note: Originally issued in Feb 1942 as a classified Nat. Defense Res.
  Council Rep.</note>
</bibl>

<bibl id="B26">
  <title><p>The Design and Implementation of {FFTW3}</p></title>
  <aug>
    <au><snm>Frigo</snm><fnm>M</fnm></au>
    <au><snm>Johnson</snm><fnm>SG</fnm></au>
  </aug>
  <source>Proceedings of the IEEE</source>
  <pubdate>2005</pubdate>
  <volume>93</volume>
  <issue>2</issue>
  <fpage>216</fpage>
  <lpage>-231</lpage>
  <note>Special issue on ``Program Generation, Optimization, and Platform
  Adaptation''</note>
</bibl>

<bibl id="B27">
  <title><p>PEP 234 -- Iterators</p></title>
  <aug>
    <au><snm>Ka Ping Yee</snm><fnm>GvR</fnm></au>
  </aug>
  <pubdate>2016</pubdate>
  <url>https://www.python.org/dev/peps/pep-0234/</url>
</bibl>

</refgrp>
} 



\clearpage
\section*{Figures}
  \begin{figure}[h!]
  \caption{\csentence{Wiener filter reconstruction}}\label{fig_wiener_filter}
          \includegraphics[width=0.97\linewidth]{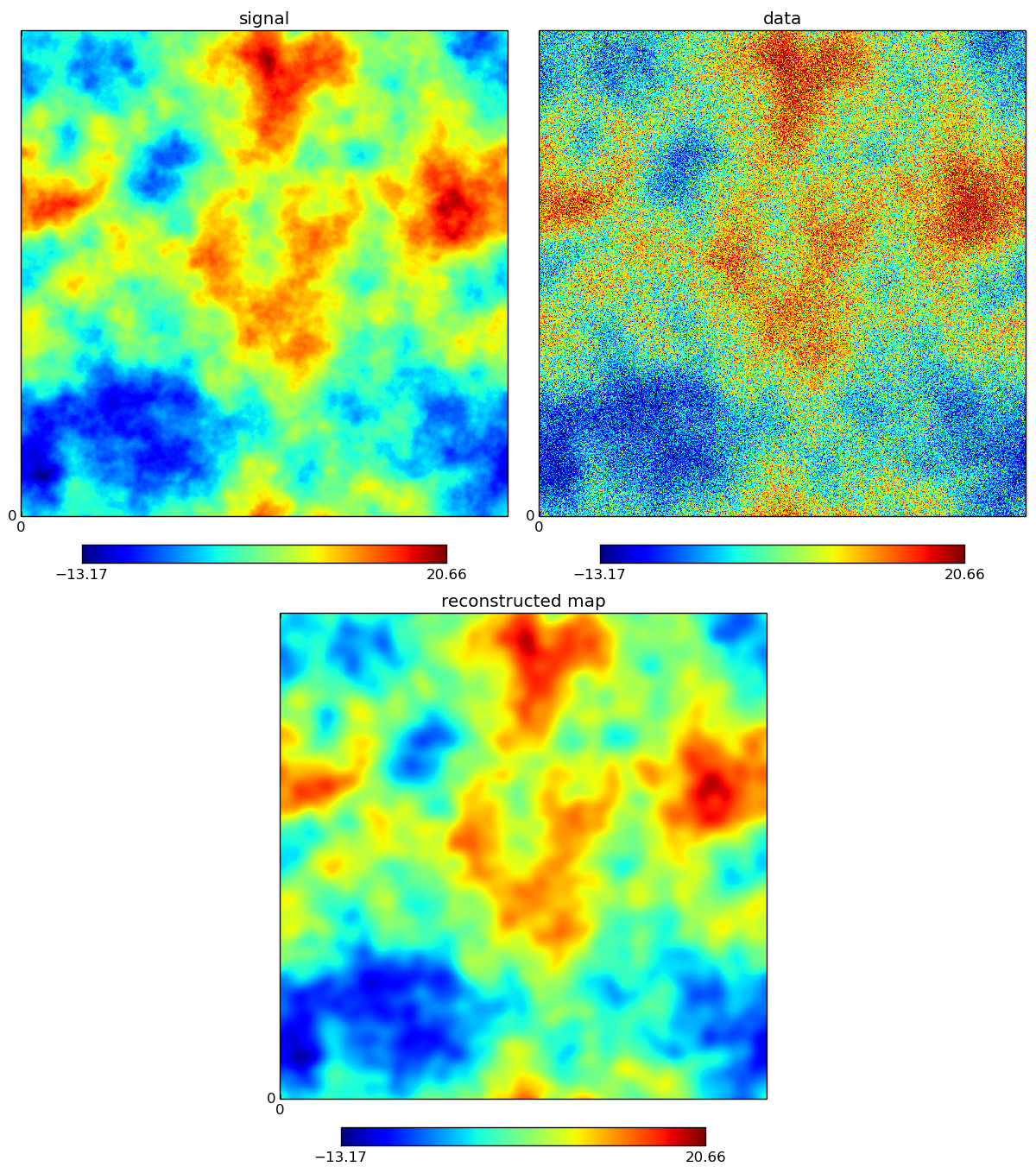}
\flushleft Top left: True signal to be reconstructed. Top right: Data which is the result of smoothing the signal and adding Gaussian noise. Bottom: reconstructed maximum of the posterior (conditional) probability density function for the signal given the data $\mathcal{P}(signal|data)$.
      \end{figure}

\clearpage
\begin{landscape}
\begin{figure}[h!]
    \caption{\csentence{Object structure}} \label{fig_object_structure}
\begin{tikzpicture}[>=stealth,thick, every node/.style={shape=rectangle,draw,rounded corners,align=left}]
\node[rectangle split, rectangle split parts=2] at (0, 0) (obj1)
    {
        obj1
        \nodepart{second}
        Attributes:\\[1ex]
        \begin{varwidth}{\linewidth}
            \begin{itemize}\setlength{\itemindent}{-15pt}
                \item \texttt{data}: \texttt{<numpy.ndarray>}
                \item \texttt{index}: 1
                \item \texttt{shape}: $(256, 256)$
                \item \texttt{distribution\_strategy}: 'equal'
                \item \texttt{distributor}: \texttt{equal\_distributor\_1}
                \item {\dots}
            \end{itemize}
        \end{varwidth}
    };

\node[rectangle split, rectangle split parts=2] at (5, 0) (obj2)
    {
        obj2
        \nodepart{second}
        Attributes:\\[1ex]
        \begin{varwidth}{\linewidth}
            \begin{itemize}\setlength{\itemindent}{-15pt}
                \item \texttt{data}: \texttt{<numpy.ndarray>}
                \item \texttt{index}: 2
                \item \texttt{shape}: $(256, 256)$
                \item \texttt{distribution\_strategy}: 'equal'
                \item \texttt{distributor}: \texttt{equal\_distributor\_1}
                \item {\dots}
            \end{itemize}
        \end{varwidth}
    };

\node[rectangle split, rectangle split parts=2] at (10, 0) (obj3)
    {
        obj3
        \nodepart{second}
        Attributes:\\[1ex]
        \begin{varwidth}{\linewidth}
            \begin{itemize}\setlength{\itemindent}{-15pt}
                \item \texttt{data}: \texttt{<numpy.ndarray>}
                \item \texttt{index}: 3
                \item \texttt{shape}: $(4, 4, 4)$
                \item \texttt{distribution\_strategy}: 'not'
                \item \texttt{distributor}: \texttt{not\_distributor\_1}
                \item {\dots}
            \end{itemize}
        \end{varwidth}
    };

\node[rectangle split, rectangle split parts=2, rectangle split part align={center, left}] at (2.5, -3.5) (distr1)
    {
        equal\_distributor\_1
        \nodepart{second}
        Attributes:\\[1ex]
        \begin{varwidth}{\linewidth}
            \begin{itemize}\setlength{\itemindent}{-15pt}
                \item \texttt{global\_shape}: $(256, 256)$
                \item \texttt{local\_shape}: $(64, 256)$
                \item {\dots}
            \end{itemize}
        \end{varwidth}
    };

\node[rectangle split, rectangle split parts=2] at (10, -3.5) (distr2)
    {
        not\_distributor\_1
        \nodepart{second}
        Attributes:\\[1ex]
        \begin{varwidth}{\linewidth}
            \begin{itemize}\setlength{\itemindent}{-15pt}
                \item \texttt{global\_shape}: $(4, 4, 4)$
                \item \texttt{local\_shape}: $(4, 4, 4)$
                \item {\dots}

            \end{itemize}
        \end{varwidth}
    };

\node[rectangle split, rectangle split parts=2] at (14, -2) (lib) {\texttt{d2o\_librarian}
        \nodepart{second}
        Attributes:\\[1ex]
        \begin{varwidth}{\linewidth}
            \begin{itemize}\setlength{\itemindent}{-15pt}
                \item \texttt{library}:
                    \begin{itemize}\setlength{\itemindent}{-28pt}
                        \item \texttt{1: obj1}
                        \item \texttt{2: obj2}
                        \item \texttt{3: obj3}
                        \item {\dots}
                    \end{itemize}

                                \end{itemize}
        \end{varwidth}
};

\node[rectangle split, rectangle split parts=2] at (0, -6) (fac) {
    \texttt{distributor\_factory}
        \nodepart{second}
        Attributes:\\[1ex]
        \begin{varwidth}{\linewidth}
            \begin{itemize}\setlength{\itemindent}{-15pt}
                \item \texttt{distributor\_store}:
                    \begin{itemize}\setlength{\itemindent}{-28pt}
                        \item \texttt{equal\_distributor\_1}
                        \item \texttt{not\_distributor\_1}
                        \item {\dots}
                    \end{itemize}

                                \end{itemize}
        \end{varwidth}
    };

\draw[*-] (obj1) |- (distr1);
\draw[*-] (obj2) |- (distr1);
\draw[*-] (obj3) -- (distr2);

\draw[*-] (fac) -| (distr1);
\draw[*-] ([yshift=-1cm]fac) -| (distr2);

\draw[-*, dashed] ([xshift=0.5cm]obj1.south) |- (lib);
\draw[dashed] ([xshift=0.5cm]obj2.south) |- ([yshift=-0.64cm, xshift=0.5cm]obj2.south);
\draw[dashed] ([xshift=0.5cm]obj3.south) |- ([yshift=-0.63cm, xshift=0.5cm]obj3.south);
\end{tikzpicture}
\flushleft Here the main object composition structure of \ddo\ is shown.
\distributeddataobjects{} are composed objects, cf.\ section \ref{sec:composed_object}.
All tasks that need information related to the distribution strategy are outsourced to a \emph{distributor}.
In this figure, three \distributeddataobjects{} are shown where \verb|obj1| and \verb|obj2| share the same distributor.
This is possible because they are essentially identical:
    they have the same global shape, datatype, and distribution strategy.
Since it is expensive to instantiate new distributors, the \distributeddataobjects{} get their instance from the \verb|distributor_factory| that takes care of caching those distributors that have already been created.
Furthermore, we illustrate the \verb|d2o_librarian| that keeps weak references to the individual \distributeddataobjects{} and assigns a unique cluster-wide identifier to them: the \verb|index|.
\end{figure}

\end{landscape}


\clearpage
\begin{landscape}
\section*{Tables}

\begin{table}[h!]
\flushleft
\caption{Overhead costs: \ddo's relative performance to numpy when scaling the array size and being run with one MPI-process.
``$100\%$'' corresponds to the case when \ddo\ is as fast as numpy.
In order to guide the eye, values $<30\%$ are printed italic, values $\geq 90\%$ are printed bold.
Please see section \ref{sec:scale_size} for discussion.}\label{table_scaling_size}
\pgfplotstabletypeset[percent cells, columns/{array size}/.style={reset styles,string type,column type={l|}}]{table_1.csv}
\end{table}
\vspace{-1em}

\begin{table}[h!]
\flushleft
\caption{Weak scaling: \ddo's relative performance to the single-process case when increasing both, the number of processes and the global array size proportionally.
The arrays used for this tests had the global shape $(n*2048, 2048)$ with n being the number of processes.
By this the local data size was fixed to $2^{22}$ elements,
    which is equal to $32~\mathrm{MiB}$.
``$100\%$'' in the table corresponds to the case were the speedup is equal to the number of processes.
Example: the $95.1\%$ for \texttt{copy\_empty} on $256$ processes correspond to a speedup-factor of $243.5$.
In order to guide the eye, values $<30\%$ are printed italic, values $\geq 90\%$ are printed bold.
Please see section \ref{sec:scale_nodes} for discussion.}\label{table_scaling_nodes}

\pgfplotstabletypeset[percent cells, columns/{process count}/.style={reset styles,string type, column type={l|}}]{table_2.csv}

\end{table}

\begin{table}[h!]
\flushleft
\caption{Strong scaling: \ddo's relative performance to a single process when increasing the number of processes while fixing the global array size to $(4096, 4096)=128~\mathrm{MiB}$.
``$100\%$'' corresponds to the case were the speedup is equal to the number of processes.
Example: the $94.4\%$ for \texttt{obj+=obj} on $256$ processes correspond to a speedup-factor of $241.7$.
In order to guide the eye, values $<30\%$ are printed italic, values $\geq 90\%$ are printed bold.
Please see section \ref{sec:strong_scaling} for discussion.}\label{table_scaling_nodes_strong}
\pgfplotstabletypeset[percent cells, columns/{processes (local size)}/.style={reset styles,string type, column type={l|}}]{table_3.csv}
\end{table}

\begin{table}[h!]
\flushleft
\caption{Strong scaling comparison: \ddo's relative performance compared to DistArray\cite{distarray} when increasing the number of processes while fixing the global array size to $(4096, 4096)=128~\mathrm{MiB}$.
``$2$'' corresponds to the case were \ddo{} is twice as fast as DistArray.
Please see section \ref{sec:comp_distarray} for discussion.}\label{table_scaling_nodes_distarray}

\pgfplotstabletypeset[nonpercent cells, columns/{processes (local size)}/.style={reset styles,string type, column type={l|}}]{table_4.csv}
\end{table}

\end{landscape}

%
%
%
%

\end{backmatter}

\appendix
\section{Advanced Usage and Functional Behavior}\label{sec:adv_usage_func_behav}
Here we discuss specifics regarding the design and functional behavior of \ddo.
We set things up by importing \verb|numpy| and \verb|d2o.distributed_data_object|:
\begin{lstlisting}[language=iPython]
    |\iin| import numpy as np
    |\iin| from d2o import distributed_data_object
\end{lstlisting}

\subsection{Distribution Strategies}\label{sec:distribution_strategies}
In order to see the effect of different distribution strategies one may run the following script using three MPI processes.
In lines \ref{listing_strat_1} and \ref{listing_strat_2}, the \verb|distribution_strategy| keyword is used for explicit specification of the strategy.\\
\begin{verbatim}
mpirun -n 3 python distribution_schemes.py
\end{verbatim}

\begin{lstlisting}[language=iPython]
    # distribution_schemes.py
    from mpi4py import MPI
    import numpy as np
    from d2o import distributed_data_object
    rank = MPI.COMM_WORLD.rank

    a = np.arange(16).reshape((4, 4))
    if rank == 0: print((rank, a))

    # use 'not', 'equal' and 'fftw'
    for strategy in ['not', 'equal', 'fftw']:
        obj = distributed_data_object(
                  a, distribution_strategy=strategy) |\label{listing_strat_1}|
        print (rank, strategy, obj.get_local_data())

    # use the 'freeform' slicer
    a += rank
    obj = distributed_data_object(
              local_data=a, distribution_strategy='freeform') |\label{listing_strat_2}|
    print (rank, 'freeform', obj.get_local_data())

    full_data = obj.get_full_data()
    if rank == 0: print (rank, 'freeform', full_data)
\end{lstlisting}
The printout in line 8 shows the \verb|a| array.
\begin{verbatim}
(0, array([[ 0,  1,  2,  3],
           [ 4,  5,  6,  7],
           [ 8,  9, 10, 11],
           [12, 13, 14, 15]]))
\end{verbatim}
The ``\verb|not|'' distribution strategy stores full copies of the data on every node:
\begin{verbatim}
(0, 'not', array([[ 0,  1,  2,  3],
                  [ 4,  5,  6,  7],
                  [ 8,  9, 10, 11],
                  [12, 13, 14, 15]]))
(1, 'not', array([[ 0,  1,  2,  3],
                  [ 4,  5,  6,  7],
                  [ 8,  9, 10, 11],
                  [12, 13, 14, 15]]))
(2, 'not', array([[ 0,  1,  2,  3],
                  [ 4,  5,  6,  7],
                  [ 8,  9, 10, 11],
                  [12, 13, 14, 15]]))
\end{verbatim}
The ``\verb|equal|'', ``\verb|fftw|'' and ``\verb|freeform|'' distribution strategies are all subtypes of the \emph{slicing} distributor that cuts the global array along its first axis.
Therefore they only differ by the lengths of their subdivisions.
The ``\verb|equal|'' scheme tries to distribute the global array as equally as possible among the processes.
If the array's size makes it necessary,
    the first processes will get an additional row.
In this example the first array axis has a length of four but there are three MPI processes;
    hence, one gets a distribution of $(2, 1, 1)$:
\begin{verbatim}
(0, 'equal', array([[0, 1, 2, 3],
                    [4, 5, 6, 7]]))
(1, 'equal', array([[ 8,  9, 10, 11]]))
(2, 'equal', array([[12, 13, 14, 15]]))
\end{verbatim}
The ``\verb|fftw|'' distribution strategy is very similar to ``\verb|equal|'' but uses functions from FFTW\cite{Frigo:1999:FFT:301618.301661}.
If the length of the first array axis is large compared to the number of processes they will practically yield the same distribution pattern but for small arrays they may differ.
For performance reasons FFTW prefers multiples of two over a uniform distribution, hence one gets $(2, 2, 0)$:
\begin{verbatim}
(0, 'fftw', array([[0, 1, 2, 3],
                   [4, 5, 6, 7]]))
(1, 'fftw', array([[ 8,  9, 10, 11],
                   [12, 13, 14, 15]]))
(2, 'fftw', array([], shape=(0, 4), dtype=int64))
\end{verbatim}
A ``\verb|freeform|'' array is built from a process-local perspective: each process gets its individual local data.
In our example, we use \verb|a+rank| as the local data arrays -- each being of shape $(4,4)$ -- during the initialization of the \distributeddataobject.
By this, a global shape of $(12, 4)$ is produced.
The local data reads:
\begin{verbatim}
(0, 'freeform', array([[ 0,  1,  2,  3],
                       [ 4,  5,  6,  7],
                       [ 8,  9, 10, 11],
                       [12, 13, 14, 15]]))
(1, 'freeform', array([[ 1,  2,  3,  4],
                       [ 5,  6,  7,  8],
                       [ 9, 10, 11, 12],
                       [13, 14, 15, 16]]))
(2, 'freeform', array([[ 2,  3,  4,  5],
                       [ 6,  7,  8,  9],
                       [10, 11, 12, 13],
                       [14, 15, 16, 17]]))
\end{verbatim}
This yields a global shape of $(12, 4)$.
In oder to consolidate the data the method \verb|obj.get_full_data()| is used,
    cf.\ section \ref{sec:advanced_indexing}.
\begin{verbatim}
(0, 'freeform', array([[ 0,  1,  2,  3],
                       [ 4,  5,  6,  7],
                       [ 8,  9, 10, 11],
                       [12, 13, 14, 15],
                       [ 1,  2,  3,  4],
                       [ 5,  6,  7,  8],
                       [ 9, 10, 11, 12],
                       [13, 14, 15, 16],
                       [ 2,  3,  4,  5],
                       [ 6,  7,  8,  9],
                       [10, 11, 12, 13],
                       [14, 15, 16, 17]]))
\end{verbatim}

\subsection{Initialization}
There are several different ways of initializing a \distributeddataobject.
In all cases its shape and data type must be specified implicitly or explicitly.
In the previous section we encountered the basic way of supplying an initial data array which then gets distributed:
\begin{lstlisting}[language=iPython]
    |\iin| a = np.arange(12).reshape((3, 4))
    |\iin| obj = distributed_data_object(a)
    # equivalent to line above
    |\iin| obj = distributed_data_object(global_data=a)
\end{lstlisting}
The initial data is interpreted as global data.
The default distribution strategy\footnote{Depending on whether \emph{pyfftw} is available or not, the \emph{equal}- or the \emph{fftw}-distribution strategy is used, respectively; cf.\ section \ref{sec:distribution_strategies}.} is a \emph{global-type} strategy,
    which means that the distributor which is constructed at initialization time derives its concrete data partitioning from the desired global shape and data type.
A more explicit example for an initialization is:
\begin{lstlisting}[language=iPython]
    |\iin| obj = distributed_data_object(global_data=a,
                                         dtype=np.complex)
\end{lstlisting}
In contrast to \verb|a|'s data type which is \emph{integer} we enforce the \distributeddataobject{} to be \emph{complex}.
Without initial data -- cf.\ \verb|np.empty| -- one may use the \verb|global_shape| keyword argument:
\begin{lstlisting}[language=iPython]
    |\iin| obj = distributed_data_object(global_shape=(2,3),
                                         dtype=np.float)
    # equivalent to line above
    |\iin| obj = distributed_data_object(global_shape=(2,3))
\end{lstlisting}
If the data type is specified neither implicitly by some initial data nor explicitly via \verb|dtype|,
    \distributeddataobject{} uses \verb|float| as a default\footnote{This mimics numpys behavior.}.
In contrast to \emph{global-type}, \emph{local-type} distribution strategies like ``\verb|freeform|'' are defined by local shape information.
The aptly named analoga to \verb|global_data| and \verb|global_shape| are \verb|local_data| and \verb|local_shape|,
    cf.\ section \ref{sec:distribution_strategies}:
\begin{lstlisting}[language=iPython]
    |\iin| obj = distributed_data_object(
                     local_data=a,
                     distribution_strategy='freeform')
\end{lstlisting}
If redundant but conflicting information is provided
    -- like integer-type initialization array vs. \verb|dtype=complex| --
    the explicit information gained from \verb|dtype| is preferred over implicit information provided by \verb|global_data|/\verb|local_data|.
On the contrary, if data is provided, explicit information from
    \verb|global_shape|/\verb|local_shape| is discarded.
In summary, \verb|dtype| takes precedence over \verb|global_data|/\verb|local_data| which in turn takes precedence over \verb|global_shape|/\verb|local_shape|.\\
Please note that besides numpy arrays, \distributeddataobjects{} are valid input for \verb|global_data|/\verb|local_data|, too.
If necessary, a redistribution of data will be performed internally.
When using \verb|global_data| this will be the case if the distribution strategies of the input and ouput \distributeddataobjects\ do not match.
When \distributeddataobjects\ are used as \verb|local_data| their full content will be concentrated on the individual processes.
This means that if one uses the same \distributeddataobject{} as \verb|local_data| in, for example, two processes,
    the resulting \distributeddataobject{} will have twice the memory footprint.

\subsection{Getting and Setting Data}\label{sec:advanced_indexing}
There exist three different methods for getting and setting a \distributeddataobject{}'s data:
\begin{itemize}
    \item \verb|get_full_data| consolidates the full data into a numpy array,
    \item \verb|set_full_data| distributes a given full-size array,
    \item \verb|get_data| extracts a part of the data and returns it packed in a new \distributeddataobject
    \item \verb|set_data| modifies parts of the \distributeddataobject{}'s data,
    \item \verb|get_local_data| returns the process' local data,
    \item \verb|set_local_data| directly modifies the process' local data.
\end{itemize}
In principle, one could mimic the behavior of \verb|set_full_data| with \verb|set_data|
    but the former is faster since there are no indexing checks involved.
\distributeddataobjects{} support large parts of numpy's indexing functionality,
    via the methods \verb|get_data| and \verb|set_data|\footnote{These are the methods getting called through Python's \verb|obj[...]=...| notation.}.
This includes simple and advanced indexing, slicing and boolean extraction.
Note that multidimensional advanced indexing is currently not supported by the slicing distributor:
    something like
    \begin{verbatim}
        obj[(np.array([[1,2], [0,1]]), np.array([[0,1], [2,3]]))]
    \end{verbatim}
\vspace{-1.3em} will throw an exception.
\begin{lstlisting}[language=iPython]
    |\iin| a = np.arange(12).reshape(3, 4)
    |\iin| obj = distributed_data_object(a)
    |\iin| obj
    |\out| <distributed_data_object>
            array([[ 0,  1,  2,  3],
                   [ 4,  5,  6,  7],
                   [ 8,  9, 10, 11]])

    # Simple indexing
    |\iin| obj[2,1]
    |\out| 9

    # Advanced indexing
    |\iin| index_tuple = (np.array([1, 1, 2, 2, 2, 2]),
                           np.array([2, 3, 0, 1, 2, 3]))
    |\iin| obj[index_tuple]
    |\out| <distributed_data_object>
            array([ 6,  7,  8,  9, 10, 11])

    # Slicing
    |\iin| obj[:, ::-2]
    |\out| <distributed_data_object>
            array([[ 3,  1],
                   [ 7,  5],
                   [11,  9]])

    # Boolean extraction
    |\iin| obj[obj>5]
    |\out| <distributed_data_object>
            array([ 6,  7,  8,  9, 10, 11])
\end{lstlisting}
All those indexing variants can also be used for setting array data, for example:
\begin{lstlisting}[language=iPython]
    |\iin| a = np.arange(12).reshape(3, 4)
    |\iin| obj = distributed_data_object(a)
    |\iin| obj[obj>5] = [11, 22, 33, 44, 55, 66]
    |\iin| obj
    |\out| <distributed_data_object>
            array([[ 0,  1,  2,  3],
                   [ 4,  5, 11, 22],
                   [33, 44, 55, 66]])
\end{lstlisting}

Allowed types for input data are scalars, tuples, lists, numpy ndarrays and \distributeddataobjects{}.
Internally the individual processes then extract the locally relevant portion of it.\\
As it is extremely costly, \ddo\ tries to avoid inter-process communication whenever possible.
Therefore, when using the \verb|get_data| method the returned data portions remain on their processes.
In case of a \distributeddataobject\ with a slicing distribution strategy the \emph{freeform distributor} is used for this, cf.\ section \ref{sec:distribution_strategies}.

\subsection{Local Keys}\label{sec:local_keys}
The distributed nature of \ddo\ adds an additional degree of freedom when getting (setting) data from (to) a \distributeddataobject.
The indexing discussed in section \ref{sec:advanced_indexing} is based on the assumption that the involved key- and data-objects are the same for every MPI node.
But in addition to that, \ddo\ allows the user to specify node-individual keys and data.
This, for example, can be useful when data stored as a \distributeddataobject\ must be piped into a software module which needs very specific portions of the data on each MPI process.
If one is able to describe those data portions as array-indexing keys -- like slices --
    then the user can do this data redistribution within a single line.
The following script -- executed by two MPI processes -- illustrates the point of local keys.
\begin{verbatim}
mpirun -n 2 python local_keys.py
\end{verbatim}

\begin{lstlisting}[language=iPython]
    # local_keys.py
    from mpi4py import MPI
    import numpy as np
    from d2o import distributed_data_object
    rank = MPI.COMM_WORLD.rank

    # initializing some data
    obj = distributed_data_object(np.arange(16)*2)

    print (rank, obj)

    # getting data using the same slice on both processes
    print (rank, obj.get_data(key=slice(None, None, 2))) |\label{listing_local_keys_1}|

    # getting data using different slices
    print (rank, obj.get_data(key=slice(None, None, 2+rank),
                              local_keys=True)) |\label{listing_local_keys_2}|

    # getting data using different distributed_data_objects
    key_tuple = (distributed_data_object([1, 3, 5, 7]),
                 distributed_data_object([2, 4, 6, 8]))
    key = key_tuple[rank]
    print (rank, obj.get_data(key=key, local_keys=True)) |\label{listing_local_keys_3}|
\end{lstlisting}
The first print statement shows the starting data: the even numbers ranging from 0 to 30:
\begin{verbatim}
(0, <distributed_data_object>
    array([ 0,  2,  4,  6,  8, 10, 12, 14]))
(1, <distributed_data_object>
    array([16, 18, 20, 22, 24, 26, 28, 30]))
\end{verbatim}
In line \ref{listing_local_keys_1} we extract every second entry from \verb|obj| using \verb|slice(None, None, 2)|.
Here, no inter-process communication is involved;
    the yielded data remains on the original node.
The output of the print statement reads:
\begin{verbatim}
(0, <distributed_data_object>
    array([ 0,  4,  8, 12]))
(1, <distributed_data_object>
    array([16, 20, 24, 28]))
\end{verbatim}
In line \ref{listing_local_keys_2} the processes ask for different slices of the global data using the keyword \verb|local_keys=True|:
process 0 requests every second element whereas process 1 requests every third element from \verb|obj|.
Now communication is required to redistribute the data and the results are stored in the individual processes.
\begin{verbatim}
(0, <distributed_data_object>
    array([ 0,  4,  8, 12, 16, 20, 24, 28]))
(1, <distributed_data_object>
    array([ 0,  6, 12, 18, 24, 30]))
\end{verbatim}
In line \ref{listing_local_keys_3} we use \distributeddataobjects{} as indexing objects.
Process 0 requests the elements at positions 1, 3, 5 and 7;
    process 1 for those at 2, 4, 6 and 8.
The peculiarity here is that the keys are not passed to \verb|obj| as a complete set of local \distributeddataobject\ instances.
In fact, the processes only hand over their very local instance of the keys.
\ddo\ is aware of this and uses the \verb|d2o_librarian| in order to reassemble them,
    cf.\ section \ref{sec:librarian}.
The output reads:
\begin{verbatim}
(0, <distributed_data_object>
    array([ 2,  6, 10, 14]))
(1, <distributed_data_object>
    array([ 4,  8, 12, 16]))
\end{verbatim}
The \verb|local_keys| keyword is also available for the \verb|set_data| method.
In this case the keys as well as the data updates will be considered local objects.
The behaviour is analogous to the one of \verb|get_data|:
    The individual processes store the locally relevant part of the \verb|to_key| using their distinct \verb|data[from_key]|.

\subsection{The d2o Librarian}\label{sec:librarian}
A \distributeddataobject\ as an abstract entity in fact consists of a set of Python objects that reside in memory of each MPI process.
Global operations on a \distributeddataobject\ necessitate that all those local instances of a \distributeddataobject\ receive the same function calls; otherwise unpredictable behavior or a deadlock could happen.
Let us discuss an illustrating example, the case of extracting a certain piece of data from a \distributeddataobject\ using slices, cf.\ section \ref{sec:advanced_indexing}.
Given a request for a slice of data, the MPI processes check which part of their data is covered by the slice, and build a new \distributeddataobject\ from that.
Thereby they communicate the size of their local data, maybe make sanity checks, and more.
If this \verb|get_data(slice(...))| function call is not made on every process of the cluster, a deadlock will occur as the `called' processes wait for the `uncalled' ones.
However, especially when using the \verb|local_keys| functionality described in section \ref{sec:local_keys}
    algorithmically one would like to work with different, i.e.\ node-individual \distributeddataobjects{} at the same time.
This raises the question: given only one local Python object instance,
    how could one make a global call on the complete \distributeddataobject\ entity it belongs to?
For this the \verb|d2o_librarian| exists.
During initialization every \distributeddataobject\ registers itself with the \verb|d2o_librarian| which returns a unique index.
Later, this index can be used to assemble the full \distributeddataobject\ from just a single local instance.
The following code illustrates the workflow.
\begin{verbatim}
mpirun -n 4 python librarian.py
\end{verbatim}
\begin{lstlisting}[language=iPython]
    # librarian.py
    from mpi4py import MPI
    import numpy as np
    from d2o import distributed_data_object, d2o_librarian

    comm = MPI.COMM_WORLD
    rank = comm.rank

    # initialize four different distributed_data_objects
    obj = distributed_data_object(np.arange(16).reshape((4,4)))
    obj_list = (obj, 2*obj, 3*obj, 4*obj)

    # every process gets its part of the respective full array
    individual_object = obj_list[rank]
    individual_index = individual_object.index
    index_list = comm.allgather(individual_index)

    for index in index_list:
        # resemble the current d2o on every node
        current_object = d2o_librarian[index]
        if rank == 0: print('Index: ' + str(index))
        # take a slice of data
        print (rank, current_object[:, 2:4].get_local_data())
\end{lstlisting}

The output reads:

\begin{verbatim}
Index: 1
(0, array([[2, 3]]))
(1, array([[6, 7]]))
(2, array([[10, 11]]))
(3, array([[14, 15]]))
Index: 2
(0, array([[4, 6]]))
(1, array([[12, 14]]))
(2, array([[20, 22]]))
(3, array([[28, 30]]))
Index: 3
(0, array([[6, 9]]))
(1, array([[18, 21]]))
(2, array([[30, 33]]))
(3, array([[42, 45]]))
Index: 4
(0, array([[ 8, 12]]))
(1, array([[24, 28]]))
(2, array([[40, 44]]))
(3, array([[56, 60]]))
\end{verbatim}

The \ddo-librarian's core-component is a \emph{weak dictionary} wherein \emph{weak references} to the local \distributeddataobject\ instances are stored.
Its peculiarity is that those weak references do not prevent Python's garbage collector from deleting the object once no regular references to it are left.
By this, the librarian can keep track of the \distributeddataobjects\ without, at the same time, being a reason to hold them in memory.

\subsection{Copy Methods}
\ddo's array copy methods were designed to avoid as much Python overhead as possible.
Nevertheless, there is a speed penalty compared to pure numpy arrays for a single process;
    cf.\ section \ref{sec:performance_scalability} for details.
This is important as binary operations like addition or multiplication of an array need a copy for returning the results.
A special feature of \ddo\ is that during a full copy one may change certain array properties such as the data type and the distribution strategy:
\begin{lstlisting}[language=iPython]
    |\iin| a = np.arange(4)
    |\iin| obj = distributed_data_object(a) # dtype == np.int
    |\iin| p = obj.copy(dtype=np.float,
                         distribution_strategy='not')
    |\iin| (p.distribution_strategy, p)
    |\out| ('not', <distributed_data_object>
                    array([ 0.,  1.,  2.,  3.]))

\end{lstlisting}
When making empty copies one can also change the global or local shape:
\begin{lstlisting}[language=iPython]
    |\iin| obj = distributed_data_object(global_shape=(4,4),
                                          dtype=np.float)
    # only the shape gets changed
    |\iin| obj.copy_empty(global_shape=(2,2))
    |\out| <distributed_data_object>
            array([[  6.90860823e-310,   9.88131292e-324],
                   [  9.88131292e-324,   1.97626258e-323]])
\end{lstlisting}

\subsection{Fast Iterators}\label{sec:fast_iterators}
A large class of problems requires iteration over the elements of an array one by one~\cite{python_iterators}.
Whenever possible, Python uses special \emph{iterators} for this in order to keep computational costs at a minimum.
A toy example is
\begin{lstlisting}[language=iPython]
    |\iin| l = [9, 8, 7, 6]
    |\iin| for item in l:
                print item
    ....:
    9
    8
    7
    6
\end{lstlisting}
Inside Python, the \verb|for| loop requests an iterator object from the list \verb|l|.
Then the loop pulls elements from this iterator until it is exhausted.
If an object is not able to return an iterator, the \verb|for| loop will extract the elements using \verb|__getitem__| over and over again.
In the case of \distributeddataobjects{} the latter would be extremely inefficient as every \verb|__getitem__| call incorporates a significant amount of communication.
In order to circumvent this, the iterators of \distributeddataobjects{} communicate the process' data in chunks that are as big as possible.
Thereby we exploit the knowledge that the array elements will be fetched one after another by the iterator.
An examination of the performance difference is done in appendix \ref{sec:perf_iter}.

\section{Iterator Performance}\label{sec:perf_iter}
As discussed in section \ref{sec:fast_iterators}, iterators are a standard tool in Python by which objects control their behavior in \verb|for| loops and list comprehensions \cite{python_iterators}.
In order to speed up the iteration process, \distributeddataobjects\ communicate their data as chunks chosen to be as big as possible.
Thereby \ddo\ builds upon the knowledge that elements will be fetched one after another by the iterator as long as further elements are requested.\footnote{This has the downside, that if the iteration was stopped prematurely, data has been communicated in vain.}
Additionally, by its custom iterator interface \ddo\ avoids that the full data consolidation logic is invoked for every entry.
Because of this, the performance gain is roughly a factor of 30 even for single-process scenarios as demonstrated in the following example:

\begin{lstlisting}[language=iPython]
    |\iin| length = 1000
    |\iin| obj = distributed_data_object(np.arange(length))

    |\iin| def using_iterators(obj):
               for i in obj:
                   pass

    |\iin| def not_using_iterators(obj):
               for j in xrange(length):
                   obj[j]

    |\iin| %timeit not_using_iterators(obj)
            10 loops, best of 3: 104 ms per loop

    |\iin| %timeit using_iterators(obj)
            100 loops, best of 3: 2.92 ms per loop
\end{lstlisting}

\end{document}